\documentclass[cameraready]{Interspeech}

\usepackage{pgfplots}

\usepackage{cite}
\usepackage{amsmath,amssymb,amsfonts}
\usepackage{algorithmic}
\usepackage{graphicx}
\usepackage{algorithm,algorithmic}
\usepackage{hyperref}
\hypersetup{hidelinks=true}
\usepackage{textcomp}

\usepackage{comment}
\usepackage{booktabs}
\usepackage[table]{xcolor} 
\usepackage{colortbl} 

\definecolor{low}{RGB}{198, 239, 206}    
\definecolor{medium}{RGB}{255, 235, 156} 
\definecolor{high}{RGB}{255, 199, 206}   

\usepackage{cite}
\usepackage{amsmath,amssymb,amsfonts}
\usepackage{algorithmic}
\usepackage{graphicx}
\usepackage{textcomp}
\usepackage{xcolor}
\usepackage{caption, subcaption}
\usepackage{amsmath,graphicx}
\usepackage{textcomp}
\usepackage{amsmath,amssymb,amsfonts}
\usepackage{mathtools}
\usepackage{amsmath,graphicx, multirow, hyperref}
\usepackage{subcaption}
\usepackage{array}     
\usepackage{booktabs}  
\usepackage{enumitem}  
\usepackage{ragged2e}  
\usepackage{dirtree}

\newcommand{\cognospeak}{CognoMemory}
\newcommand{\cognospeakOLD}{CognoSpeak}
\newcommand{\dementiabank}{DementiaBank}

\newcommand{\roberta}{RoBERTa}
\newcommand{\distilbert}{DistilBERT}

\newcommand{\compare}{ComParE}
\newcommand{\adress}{ADReSS}

\newcommand{\Whisper}{Whisper}
\newcommand{\WV}{Wav2Vec 2.0}

\title{PROCESS-2: A Benchmark Speech Corpus for Early Cognitive Impairment Detection}

\author[affiliation={1}]{Madhurananda}{Pahar}
\author[affiliation={2}]{Caitlin H.}{Illingworth}
\author[affiliation={1}]{Bahman}{Mirheidari}
\author[affiliation={1}]{Hend}{Elghazaly}
\author[affiliation={1}]{Fritz}{Peters}
\author[affiliation={2}]{Sophie}{Young}
\author[affiliation={1}]{Wing-Zin}{Leung}
\author[affiliation={2}]{Labhpreet}{Kaur}
\author[affiliation={2}]{Daniel}{Blackburn}
\author[affiliation={1}]{Heidi}{Christensen}

\address{
    $^1$ School of Computer Science, University of Sheffield, Sheffield, S1 4DP, UK \\
    $^2$ Sheffield Institute for Translational Neuroscience (SITraN), University of Sheffield, Sheffield, S10 2HQ, UK
}

\email{\{m.pahar, chillingworth1, b.mirheidari, helghazaly1, fpeters3, syoung6, wleung5, lkaur2, d.blackburn, heidi.christensen\}@sheffield.ac.uk}

\usepackage{comment}

\begin{document}

\maketitle

\begin{abstract}
    
    Speech-based analysis offers a scalable and non-invasive approach for detecting cognitive decline, yet progress has been constrained by the limited availability of clinically validated datasets collected under realistic conditions. We introduce PROCESS-2, a large-scale speech dataset designed to support research on automatic assessment of cognitive impairment from spontaneous and task-oriented speech. The dataset comprises recordings from 200 healthy controls, 150 mild cognitive impairment, and 50 dementia diagnoses collected using the \cognospeak{} digital assessment platform. Each participant completed a single assessment session, including picture description and verbal fluency tasks, accompanied by manually verified transcripts and participant-level metadata.
    PROCESS-2 contains approximately 21 hours of speech audio with predefined train/test partitions. Comprehensive technical validation evaluated demographic balance, clinical consistency, recording stability, embedding-space structure, and reproducible baseline modelling performance, demonstrating clinically meaningful group separation and stable performance across modelling approaches while preserving real-world conversational variability.
    PROCESS-2 is released under controlled access via Hugging Face to enable responsible reuse while protecting participant privacy,  providing a reproducible benchmark resource for speech-based cognitive assessment research.

\end{abstract}


\section{Background \& Summary}


\subsection{Scientific Context}

Neurodegenerative disorders associated with cognitive decline, including mild cognitive impairment (MCI) and dementia, represent a major and rapidly expanding global health challenge driven by demographic ageing \cite{davis2018estimating}. 
Cognitive deterioration affects memory, executive function, and language production, with early changes frequently emerging in spontaneous speech long before functional impairment becomes clinically evident \cite{rosenberg2013association, knopman2003essentials, thabtah2020correlation}. 
Speech production engages multiple cognitive systems simultaneously, including semantic retrieval, working memory, attention, and executive control, making spontaneous speech an attractive non-invasive biomarker for early detection and monitoring of cognitive decline \cite{prestia2013prediction, hendrie1998epidemiology}. 
Early identification is clinically important because timely diagnosis enables intervention planning, patient support, and longitudinal monitoring that may delay progression and reduce healthcare burden \cite{shi2023speech, mckhann2011diagnosis}. 
However, current diagnostic pathways rely heavily on specialist clinical evaluation, neuroimaging, and invasive biomarker testing, which are costly, time-consuming, and difficult to scale for population-level screening \cite{yang2022deep}. 
As a result, a substantial proportion of individuals experiencing early cognitive decline remain undiagnosed worldwide, placing increasing strain on healthcare systems managing ageing populations \cite{mckhann2011diagnosis}.
There is, therefore, an increased necessity for remote, smart technologies to support healthcare services to deliver timely and accurate diagnosis \cite{gauthier2021world}.

\begin{table*}[!ht]
\centering
\footnotesize
\renewcommand\arraystretch{1.5}
\caption{Comparison of major speech datasets for cognitive decline research and the proposed PROCESS-2 dataset.}
\label{tab:dataset_comparison}

\newlist{tabitemize}{itemize}{1}
\setlist[tabitemize]{label=\textbullet, leftmargin=*, nosep, before=\vspace{5pt}, after=\vspace{3pt}}

\begin{tabular}{>{\centering\arraybackslash}m{2cm} >{\centering\arraybackslash}m{0.4cm} >{\centering\arraybackslash}m{1.1cm} >{\centering\arraybackslash}m{1.8cm} >{\centering\arraybackslash}m{2.4cm} m{3.5cm} m{3.5cm}}
\hline
\textbf{Dataset} & \textbf{Year} & \textbf{Language} & \textbf{Setting} & \textbf{N (Split)} & \textbf{Main Strengths} & \textbf{Key Limitations} \\
\hline

DementiaBank Pitt Corpus \cite{becker1994natural} 
& 1994
& English 
& Controlled clinical recordings
& 397 (104 HC / 208 AD / 85 Others)
& \begin{tabitemize}
    \item Foundational benchmark
    \item Longitudinal data available
    \item Standardised tasks: CTD, story recall, verbal fluency, and sentence repetition
  \end{tabitemize} 
& \begin{tabitemize}
    \item Older, low-quality audio
    \item Low ecological validity
    \item Limited to binary classes (AD vs. HC)
  \end{tabitemize} \\ \hline

I-CONECT / Dodge \cite{dodge2014characteristics}
& 2014
& English 
& Remote webcam interviews 
& 83-320 (42 MCI / 41 HC pilot)
& \begin{tabitemize}
    \item Longitudinal conversational data
    \item High ecological validity (home-based)
  \end{tabitemize} 
& \begin{tabitemize}
    \item Significant acoustic noise (fans, etc.)
    \item High participant dropout risk
  \end{tabitemize} \\ \hline

Dem@Care \cite{karakostas2016care}
& 2016
& Greek 
& Multi-sensor lab/home 
& $\sim$50 (22 AD / 28 HC)
& \begin{tabitemize}
    \item Multi-sensor (audio, video, depth)
    \item Focus on Activities of Daily Living (ADL)
  \end{tabitemize} 
& \begin{tabitemize}
    \item Very small sample size
    \item High technical complexity (syncing)
  \end{tabitemize} \\ \hline

\adress{} Challenge \cite{luz2020alzheimer}
& 2020
& English 
& Controlled clinical recordings 
& 156 (78 AD / 78 HC) 
& \begin{tabitemize}
    \item Balanced age/gender sets
    \item Standardised ML splits
  \end{tabitemize} 
& \begin{tabitemize}
    \item Single task (CTD)
    \item Small cohort size
  \end{tabitemize} \\ \hline

ADReSSo \cite{luz2021detecting}
& 2021
& English 
& Controlled clinical recordings 
& 237 (Train: 83 AD / 83 HC; Test: 35 AD / 36 HC)
& \begin{tabitemize}
    \item Acoustic-only focus
    \item Standardised ML splits
  \end{tabitemize} 
& \begin{tabitemize}
    \item Task-restricted (CTD \& SFT) protocols
    \item Limited acoustic diversity
  \end{tabitemize} \\ \hline

NCMMSC2021 \cite{ying2023multimodal}
& 2021
& Chinese 
& Semi-structured sessions
& 124 (26 AD / 54 MCI / 44 HC)
& \begin{tabitemize}
    \item Tonal language features
    \item Natural conversation
  \end{tabitemize} 
& \begin{tabitemize}
    \item Limited public access
    \item Significant class imbalance
  \end{tabitemize} \\ \hline

Japanese Corpus \cite{igarashi2022cognitive}
& 2022
& Japanese 
& Quiet lab environment 
& 60 (12 AD / 21 MCI / 27 HC)
& \begin{tabitemize}
    \item Episodic/emotional tasks
    \item Inclusion of MCI cases
  \end{tabitemize} 
& \begin{tabitemize}
    \item Very small sample size
    \item Difficult to scale
  \end{tabitemize} \\ \hline

Talk2Me / FVS \cite{amini2022automated}
& 2022
& English 
& Structured interviews 
& 1000+ (Varies; e.g., $\sim$100 per class)
& \begin{tabitemize}
    \item Large-scale MCI cohort
    \item High statistical power
  \end{tabitemize} 
& \begin{tabitemize}
    \item Restricted data sharing
    \item Limited multimodal access
  \end{tabitemize} \\ \hline

ADReSS-M \cite{luz2023multilingual}
& 2023
& English / Greek 
& Standard interviews  
& 291 (EN: 83 AD / 83 HC; GR: 22 AD / 24 HC)
& \begin{tabitemize}
    \item Multilingual evaluation
    \item High-fidelity recordings
  \end{tabitemize} 
& \begin{tabitemize}
    \item Imbalanced language subsets
    \item Highly controlled setting
  \end{tabitemize} \\ \hline

ADReFV \cite{xu2023adrefv}
& 2023
& Chinese 
& HCI-based facial tasks 
& 150 (75 AD / 75 HC)
& \begin{tabitemize}
    \item Unique facial/micro-expression data
    \item Interactive execution tasks
  \end{tabitemize} 
& \begin{tabitemize}
    \item Single modality (facial video)
    \item Lab-based/less naturalistic
  \end{tabitemize} \\ \hline

TAUKADIAL \cite{barrera2024interspeech}
& 2024
& English / Chinese 
& Controlled picture tasks 
& 169 (84 MCI / 85 HC)
& \begin{tabitemize}
    \item Standardised multilingual cohort
    \item Focus on early-stage MCI
  \end{tabitemize} 
& \begin{tabitemize}
    \item Task-restricted (3 images only)
    \item Binary focus (MCI vs HC)
  \end{tabitemize} \\ \hline

PROCESS Challenge \cite{tao2025PROCESS}
& 2025
& English 
& Remote real-world (home, clinic, community) 
& 197 (20 AD / 74 MCI / 103 HC)
& \begin{tabitemize}
    \item Remote cohort with CTD, SFT \& PFT tasks
    \item Real-world noise diversity
  \end{tabitemize} 
& \begin{tabitemize}
    \item Not released publicly
    \item Imbalanced
  \end{tabitemize} \\ \hline

CogPic \cite{wu2026cogpic} 
& 2026
& Chinese 
& Semi-structured sessions
& 574 (140 AD / 256 MCI / 178 HC)
& \begin{tabitemize}
    \item Multimodal (audio, video, text)
    \item A large corpus and metadata with education and MoCA
  \end{tabitemize} 
& \begin{tabitemize}
    \item CTD task Only
  \end{tabitemize} \\ \hline

\textbf{PROCESS-2 (This study)}
& \textbf{2026}
& \textbf{English (UK)} 
& \textbf{Remote real-world}
& \textbf{400 (50 AD / 150 MCI / 200 HC)}
& \begin{tabitemize}
    \item Ecologically valid speech
    \item Multiple tasks: CTD, SFT \& PFT
    \item Large-scale remote cohort
    \item Real-world noise diversity
  \end{tabitemize} 
& \begin{tabitemize}
    \item Limited participants (174) with MMSE
    \item British English only 
  \end{tabitemize} \\ \hline

\end{tabular}
\vspace{-15pt}
\end{table*}

\subsection{Current Limitations}

Advances in computational speech analysis and machine learning have demonstrated strong potential for detecting dementia-related cognitive changes from linguistic and acoustic features, motivating the development of scalable digital health assessments for remote monitoring \cite{pan2021using}. 
Despite rapid methodological progress, translation toward clinically deployable speech biomarkers remains constrained by limitations in available datasets, which are collected in a real-world environment and rarely achieve a balance between cohort scale, task diversity, multimodal availability, and clinically grounded diagnostic annotation.

Table~\ref{tab:dataset_comparison} presents a chronological overview of major speech datasets for cognitive decline research. Early foundational resources, most notably the DementiaBank Pitt Corpus (1994) \cite{becker1994natural}, established speech as a viable diagnostic signal and remain widely used benchmarking datasets. However, these recordings were collected primarily in controlled clinical environments using highly standardised elicitation protocols, resulting in limited ecological validity and reduced generalisability to real-world deployment settings. Furthermore, early datasets largely focused on binary diagnostic distinctions between Alzheimer’s disease and healthy controls (HC), providing limited representation of intermediate cognitive states such as MCI.

Mid-generation datasets began addressing ecological or real-world validity through remote and multimodal data collection. Studies such as I-CONECT (2014) \cite{dodge2014characteristics} introduced longitudinal home-based conversational assessments using remote communication technologies, while Dem@Care (2016) \cite{karakostas2016care} explored multimodal monitoring of cognitive decline through integrated audio, video, and behavioural sensing in laboratory and home environments. Although these initiatives represented important steps toward real-world assessment, they were typically constrained by relatively small cohort sizes, technological complexity, or limited public data accessibility, restricting large-scale, reproducible machine learning research.


Recent datasets (2020–2026) have expanded linguistic diversity and experimental scope. 
Further benchmark initiatives, including derivatives of \dementiabank{} such as \adress{} (2020) and ADReSSo (2021) \cite{luz2020alzheimer,luz2021detecting}, improved methodological comparability by introducing balanced cohorts and reproducible evaluation protocols. These datasets standardised evaluation procedures and accelerated methodological development within the field. 
Large cohort initiatives derived from structured interview studies, including Talk2Me (2022) or Framingham Voice Study (FVS) \cite{amini2022automated}, provide improved statistical power and enhanced MCI representation but often impose data-sharing restrictions that limit reproducible benchmarking. 

Multilingual and international resources such as NCMMSC2021 (2021) \cite{ying2023multimodal}, the Japanese cognitive assessment corpus (2022) \cite{igarashi2022cognitive}, ADReSS-M (2023) \cite{luz2023multilingual}, TAUKADIAL (2024) \cite{barrera2024interspeech}, and large multimodal collections, including CogPic (2026) \cite{wu2026cogpic} introduce broader language coverage and richer metadata. These efforts represent important progress toward cross-linguistic dementia assessment; however, many remain constrained by single-task paradigms or controlled recording conditions that limit ecological realism.



The PROCESS Grand Challenge dataset (2025) \cite{tao2025PROCESS}, derived from early \cognospeak{} deployments \cite{pahar2025cognospeak, pahar2025mutlimodalfusion, young2025can, pahar2025cognospeakWiley, illingworth2025developing, pahar2026can}, represents a step toward real-world remote assessment, although only a restricted subset of recordings and metadata was publicly released for benchmarking \cite{chi2025predicting, qian2025dust, zafar2025multi, zhang2025cognitive, gao2025leveraging, thallinger2025multi, illaste2025taltech}.

Across three decades of dataset development (1994–2026), several recurring limitations persist: (i) reliance on controlled or semi-structured recording environments that only partially reflect real-world telehealth deployment, (ii) dependence on narrowly defined elicitation paradigms capturing limited aspects of cognition, (iii) insufficient coverage of multiple cognitive domains, (iv) restricted accessibility of large clinically annotated cohorts, and (v) limited availability of openly shareable resources suitable for reproducible machine learning research.

These observations motivate the development of datasets that integrate large-scale participation, clinically validated diagnostic labels, multimodal setup, more diverse cognitive elicitation tasks, and ecologically valid remote acquisition across heterogeneous real-world environments, as addressed by the proposed PROCESS-2 dataset.

\subsection{Dataset contribution}


To address the limitations identified across existing resources, we introduce PROCESS-2, an extension of ``The Prediction and Recognition Of Cognitive declinE through Spontaneous Speech (PROCESS)'' Signal Processing Grand Challenge \cite{tao2025PROCESS} and a large-scale dataset of conversational speech for remote cognitive assessment collected using the \cognospeak{}, (formerly \cognospeakOLD{} \cite{pahar2025cognospeak}) automatic assessment platform. 
PROCESS-2 aims to unify ecological validity, clinically grounded diagnostic annotation, and reproducible benchmarking by integrating realistic, naturalistic, and clinically representative data within a single acquisition framework.

The dataset comprises recordings from 400 older adults recruited across the United Kingdom (UK), spanning 50 dementia, 150 MCI, and 200 cognitively healthy control participants. All data were acquired remotely through semi-structured human–computer conversational interactions conducted in real-world environments, including participants’ homes, community locations, and clinical settings. Unlike controlled environment-based datasets, recordings capture natural variability arising from heterogeneous consumer devices, background noise conditions, and spontaneous conversational behaviour, thereby reflecting realistic telehealth deployment scenarios.

PROCESS-2 incorporates three complementary speech elicitation paradigms targeting distinct cognitive and linguistic processes: SFT, PFT and CTD tasks. 
The inclusion of both structured and cognitively-demanding tasks enables investigation of lexical retrieval, executive function, semantic organisation, and discourse-level language production within a single dataset. Predefined training (80\%) and held-out test (20\%) partitions are provided to support reproducible machine learning evaluation and standardised benchmarking.

By integrating spontaneous speech recordings collected conversationally, diagnostic labels, demographic metadata, and available cognitive screening measures,
PROCESS-2 provides an ecologically valid resource for developing and evaluating automated speech-based biomarkers of early cognitive decline.

\subsection{Dataset overview}


The PROCESS-2 release provides a structured multimodal dataset comprising speech recordings and associated metadata collected through a fully remote assessment workflow. The dataset contains audio recordings from 400 participants completing three speech elicitation tasks, resulting in 1200 organised collections of task-specific audio recordings in total.

All speech data is distributed as standard waveform (.wav) audio files accompanied by aligned textual transcripts and structured metadata tables. 
Metadata includes participant demographic information such as age and gender, diagnostic category (Dementia, MCI, and HC), and available cognitive assessment scores, such as Mini-Mental State Examination (MMSE) measurements for a subset of participants. This organisation enables transparent linkage between raw recordings, linguistic content, and clinical annotations.

Recordings were obtained using participants’ own consumer devices in natural home environments, intentionally preserving acoustic variability characteristic of remote digital health deployment. Rather than minimising environmental variation, PROCESS-2 captures realistic signal diversity, allowing researchers to evaluate the robustness of speech-based models under real-world operating conditions.

The dataset supports a broad range of secondary research applications, including automated dementia screening, multimodal speech biomarker discovery, robustness analysis under heterogeneous recording conditions, and benchmarking of machine learning systems for cognitive assessment. By providing standardised data structure, predefined evaluation splits, manual transcriptions and clinically informed diagnostic labels, PROCESS-2 enables reproducible investigation of speech-based indicators of cognitive decline at scale.


\begin{figure*}[!ht]
  \centering 
  \includegraphics[width=\linewidth]{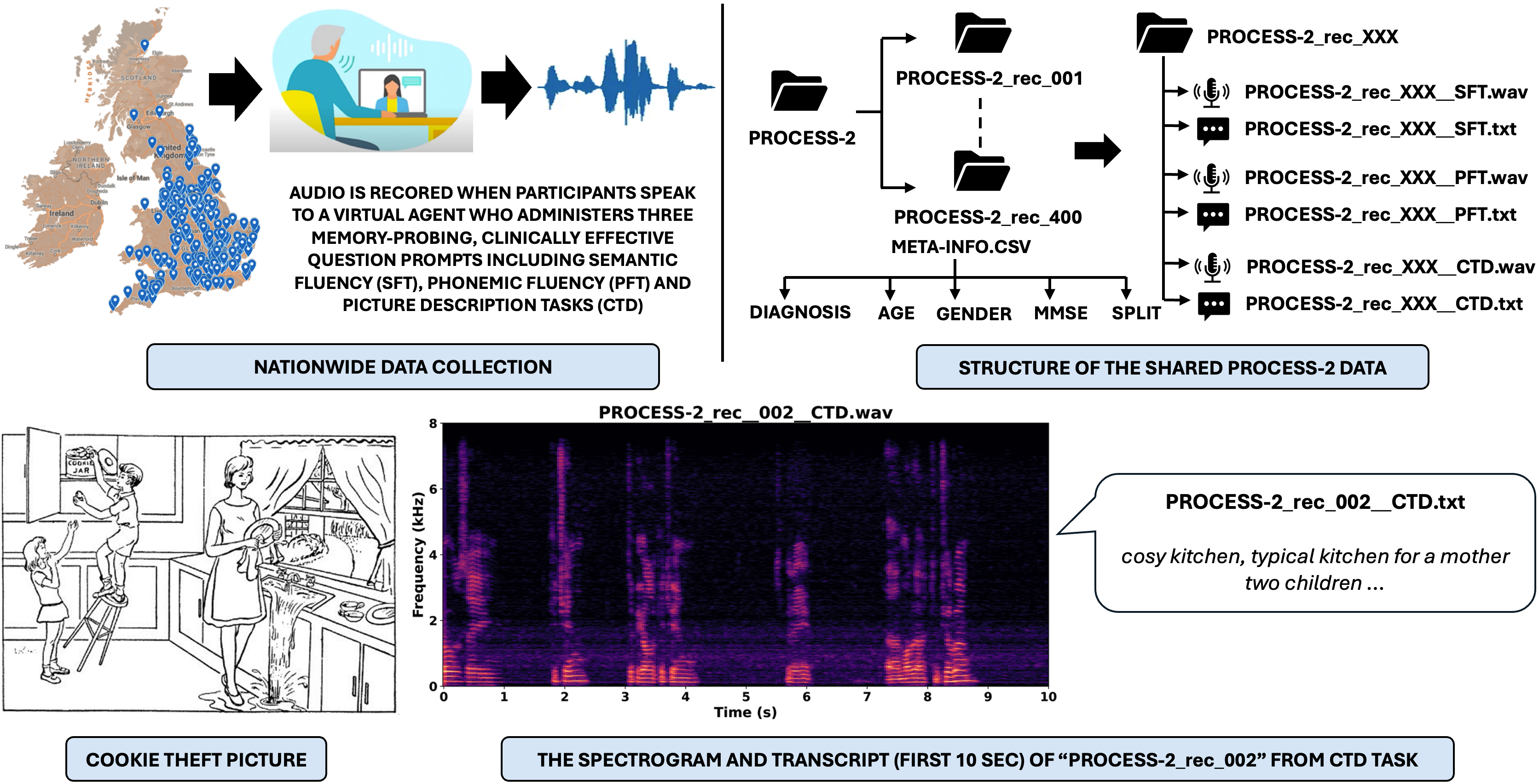} 
  \caption{\textbf{Overview of the PROCESS-2 dataset collection pipeline and shared data structure:} Speech recordings were collected nationwide using the \cognospeak{} virtual assessment platform, where participants interacted with a conversational agent administering clinically validated cognitive tasks, including semantic fluency (SFT), phonemic fluency (PFT), and Cookie Theft picture description (CTD). Audio responses were recorded during natural speech interaction and organised into participant-level directories containing task-specific waveform recordings and manually verified transcripts. Each recording is accompanied by metadata describing diagnosis, demographic variables, cognitive scores (MMSE), and predefined dataset splits for reproducible experimentation. The lower panel illustrates an example CTD task, together with the corresponding speech spectrogram and transcript excerpt from a representative recording. The figure summarises the end-to-end PROCESS-2 workflow from nationwide data acquisition to the structured research-ready dataset released to the community. } 
  \label{fig:summary}
  \vspace{-10pt}
\end{figure*}

\section{Methods}


The PROCESS-2 dataset was generated through a standardised remote assessment pipeline designed to capture spontaneous conversational speech and associated clinical metadata under real-world conditions. Data acquisition was conducted using the \cognospeak{} digital cognitive assessment platform, which supports automated recruitment, electronic consent, cognitive screening, and speech recording through a browser-based interface accessible from participants’ personal devices.

Dataset creation followed a reproducible workflow comprising participant recruitment and diagnostic verification, remote administration of speech assessment tasks, automated multimedia recording, transcription and annotation, and structured data curation into a unified repository. All participants completed identical assessment protocols delivered through the same platform infrastructure, ensuring consistent task presentation, timing, and recording procedures despite heterogeneous recording environments.

Ethical approval was obtained prior to data collection, and all participants provided informed consent permitting research participation and controlled data sharing. 
The study was conducted in accordance with the Declaration of Helsinki and ethical guidelines for research involving human participants. Ethical approval for data collection was granted by the NRES Committee South West–Central Bristol (REC number 16/LO/0737).
The following sections describe participant recruitment, clinical characterisation, platform architecture, speech elicitation procedures, and preprocessing steps required for independent replication of the PROCESS-2 dataset.

\subsection{Participants recruitment}

Participants were recruited through collaborating clinical services and research registries using the \cognospeak{} remote assessment platform to capture a representative spectrum of cognitive function and real-world assessment conditions.
Recruitment pathways included National Health Service (NHS) memory services, primary and secondary care referrals, and research volunteer registries such as Join Dementia Research (JDR) and Great Minds.  
Individuals recruited through the NHS memory services were undergoing routine clinical evaluation for suspected cognitive impairment at the time of recruitment or had recently received a clinical diagnosis. NHS recruitment sites covered multiple UK localities such as Bradford, Humber, London, Newcastle, Manchester, Barnsley, Sheffield, York, Leeds, Doncaster, Bristol, and Southampton. 
Additional participants
recruited via research volunteer registries typically had existing clinical diagnoses established prior to study participation or were enrolled as cognitively healthy volunteers. In the case that a participant disclosed a diagnosis of dementia or MCI, researchers from the University of Sheffield aided in their recruitment and conducted an additional traditional memory assessment. 
Recruitment across both pathways ensured the inclusion of participants assessed within routine healthcare services, 
thereby increasing ecological validity and demographic diversity while maintaining clinically verified diagnostic labels.

Inclusion and exclusion criteria were defined within the project clinical protocol and applied consistently across recruitment pathways (Table~\ref{tab:inclusion_exclusion}).
Hearing impairment was not an exclusion criterion provided participants were able to hear assessment prompts, including through hearing aids when required. A history of stroke or psychiatric comorbidity was recorded, but did not constitute exclusion unless it affected the capacity to consent. Medication use was not collected within the current dataset release.
Diagnoses were assigned according to the standard clinical memory clinic 
procedures using multidisciplinary evaluation, including clinical 
history, cognitive assessment, and clinician judgement, independent 
of any computational speech analyses.
The cohort primarily reflects the UK English-speaking memory clinic 
referrals and therefore may not fully represent linguistic, cultural, or healthcare-system diversity present in other regions.

\begin{table*}[t]
\centering
\caption{Participant inclusion and exclusion criteria for PROCESS-2 recruitment through \cognospeak{} platform.}
\label{tab:inclusion_exclusion}
\begin{tabular}{p{7.5cm} p{7.5cm}}
\toprule
\textbf{Inclusion Criteria} & \textbf{Exclusion Criteria} \\
\midrule
\begin{itemize}[nosep, leftmargin=*]
    \item Patients referred to memory clinics with suspected cognitive impairment (e.g., Alzheimer’s Disease, Dementia with Lewy Bodies, Parkinson’s Disease Dementia, Frontotemporal Dementia, or Functional Cognitive Disorder).
    \item Capacity to provide informed consent.
    \item Ability to engage with conversational assessment tasks.
\end{itemize} 
& 
\begin{itemize}[nosep, leftmargin=*]
    \item Lack of capacity to provide informed consent.
    \item Insufficient English comprehension for consent or assessment procedures.
    \item Severe speech impairment (e.g., profound dysphasia) preventing verbal participation.
    \item Severe motor impairment preventing interaction even with caregiver assistance.
\end{itemize} \\
\bottomrule
\end{tabular}
\vspace{-5pt}
\end{table*}

\subsection{Diagnostic procedures and Clinical metadata collection}

All participants living with dementia or MCI received diagnostic classification through qualified clinicians prior to or during recruitment. Individuals recruited through NHS clinical pathways underwent standard diagnostic evaluation consistent with National Institute for Health and Care Excellence (NICE) dementia guidelines \cite{NICE2018Dementia}. Diagnostic assessment typically included clinical history, cognitive testing, blood investigations, and structural brain imaging.
Participants recruited via research registries that disclosed a previously established clinical diagnosis also completed a researcher-administered Montreal Cognitive Assessment (MoCA) \cite{nasreddine2005montreal} via an online video call to support diagnostic categorisation. 
Clinical diagnoses were made without using any results from the computational speech analyses. The clinicians responsible for diagnosis did not have access to model outputs or research findings during the diagnostic process.
For all patient participants, cognitive assessment scores were required to be obtained within three months of the \cognospeak{} assessment. Cognitive evaluations were conducted by healthcare professionals or researchers either during clinical appointments, telephone consultations, or remote video assessments, depending on the recruitment pathway.

\subsection{\cognospeak{} platform}

Data were collected using the \cognospeak{} digital cognitive assessment platform, a browser-based system designed for automated remote evaluation of cognitive function through conversational speech interaction. The platform integrates participant onboarding, electronic consent, task delivery, automated conversational prompting, and multimedia recording within a unified interface.

Participants accessed the system remotely using personal consumer devices, including laptops and tablet computers running Windows, macOS, or iOS operating systems. Assessments were conducted through the Google Chrome browser to ensure compatibility with Web Real-Time Communication (WebRTC)-based audio and video capture.

Speech data were elicited through interaction with a virtual conversational agent developed with input from clinicians and computational linguists \cite{pahar2025cognospeak}. 
Participants selected one of four virtual agents designed to represent diverse ethnicities and age groups to promote engagement and communication comfort. The assessment included structured questioning targeting memory recall, speech fluency, cognitive functioning, reading ability, and picture description tasks commonly used in cognitive decline assessment.

The virtual conversational agent delivered all instructions and prompts, ensuring identical task presentation, wording, and timing constraints across participants, independent of geographic location or clinical supervision.
Participants completed a single assessment session comprising all conversational prompts and cognitive speech tasks. Tasks were not repeated within sessions.

\subsection{Speech elicitation tasks}

\subsubsection{Semantic Fluency Task (SFT)}

In the Semantic Fluency Task, participants were instructed by the virtual assistant to produce as many words as possible belonging to a specified semantic category (animals) within a fixed duration of one minute. The task probes semantic memory retrieval, lexical access, and executive search processes commonly affected in early cognitive decline \cite{vaughan2018semantic, henry2004verbal, olmos2023phonological}.
Audio and video responses were recorded continuously during task execution. 

\subsubsection{Phonemic Fluency Task (PFT)}

The Phonemic Fluency Task required participants to generate words beginning with a specified letter ``P" within a one-minute time limit. This task primarily assesses executive control, phonological retrieval, and cognitive flexibility \cite{steiner2008phonemic}. Standardised instructions were delivered automatically by the \cognospeak{} conversational agent prior to recording onset. 

\subsubsection{Cookie Theft Description (CTD)}

During the CTD task, participants were asked to describe a complex visual scene presented on screen. Unlike fluency tasks, no strict time limit was imposed, allowing natural spontaneous speech production, where
recordings typically ranged between approximately 60 and 75 seconds (Table~\ref{tab:audio_overview}). 
The task captures spontaneous narrative speech, discourse organisation, and pragmatic language abilities \cite{forbes2005detecting}.


\subsection{Data Acquisition and Processing}
Raw multimedia recordings were automatically uploaded from participant devices to a secure cloud infrastructure hosted via Google Firebase services. Data were subsequently downloaded to a dedicated high-performance workstation for processing and curation.

Processing was performed on a high-performance workstation equipped with an AMD EPYC CPU, 188 GB RAM, and four NVIDIA RTX 4090 GPUs.
Web-based sessions produced audio recordings encoded alongside video streams (audio: WAV; video: WEBM), while iPad/iOS devices generated M4A audio and MOV video files. 
All source media were transcoded to .wav format using FFmpeg, maintaining a constant bitrate of 128 kbps.
All audio recordings were converted to mono format, resampled from 44.1~kHz to 16~kHz, and normalised to a target loudness of $-23$~LUFS following the EBU~R128 broadcast loudness recommendation. These steps ensured consistency across heterogeneous recording devices while preserving natural speech characteristics \cite{ebu2011loudness}. 
Collected MoCA scores were subsequently converted to Mini-Mental Status Examination (MMSE) \cite{fasnacht2023conversion}.

\subsection{Data Curation}
\label{subsec:data-curation}

Recordings were manually reviewed to ensure completeness, and incomplete responses to time-limited tasks were truncated to the portion containing participant speech.

Transcripts were generated manually by multiple annotators including professional transcribers and study authors. Some transcripts include disfluencies, speaker identifiers (e.g., ``Pat:'', ``Oth:'') and pause annotations (e.g., ``(2 seconds)''). Transcript files were preserved in their original form without post-hoc modification to maintain fidelity to the transcription process.
No additional linguistic annotation schema was imposed during dataset release. Researchers may therefore apply task-specific annotation protocols depending on downstream analytical objectives.

Some assessments were conducted in clinical environments with assistance from clinicians or accompanying individuals when required. To preserve ecological validity and reflect real-world deployment conditions, conversational contributions from assisting speakers were retained within recordings when present.


All recordings were pseudonymised prior to release using anonymised participant identifiers, and no personally identifiable information was retained within filenames, transcripts, or metadata tables.
The curated dataset was organised into a participant-level hierarchical repository described in detail in Section~\ref{sec:datarecords}.

\subsection{Quality Control and reproducibility statement}

Recordings were obtained in participants’ home environments using heterogeneous consumer hardware; however, multiple procedural and software-level controls were implemented to ensure consistent acquisition. All assessments were delivered through a single web-based \cognospeak{} interface providing identical task instructions, interaction flow, and timing constraints across participants. Automated task administration eliminated examiner-related variability, while fixed recording parameters and supported web browser access enforced uniform media encoding and sampling configurations independent of device type.
Post-acquisition preprocessing harmonised recordings through format conversion, resampling, channel normalisation, and loudness standardisation. These procedures minimised variability attributable to recording conditions while intentionally preserving ecologically valid acoustic variability representative of real-world deployment environments.
All recordings and transcripts were manually reviewed to remove 
personal identifiers, including names, locations, and sensitive 
personal references prior to dataset release.
Together, these procedures define PROCESS-2 as a cross-sectional clinical speech dataset reflecting single-session real-world cognitive assessments rather than controlled laboratory recordings.

All stages of dataset creation, including task administration, recording procedures, preprocessing, transcription, and data organisation, were defined through fixed protocols and automated platform delivery. The combination of standardised virtual-agent interaction, predefined dataset partitions, consistent file naming conventions, and publicly documented preprocessing procedures enables independent researchers to reproduce the PROCESS-2 dataset structure and experimental workflows without access to proprietary infrastructure.

\section{Data Records}
\label{sec:datarecords}

The PROCESS-2 dataset is released as a structured multimodal resource comprising speech recordings, manual transcripts, and participant-level metadata. The dataset is organised within a parent directory (\texttt{PROCESS-2}) containing one subdirectory per participant.

\subsection{Directory Structure}

Each participant directory follows a consistent naming convention:


\dirtree{%
.1 PROCESS-2/.
.2 PROCESS-2\_rec\_\_XXX/.
.3 PROCESS-2\_rec\_\_XXX\_\_SFT.wav.
.3 PROCESS-2\_rec\_\_XXX\_\_SFT.txt.
.3 PROCESS-2\_rec\_\_XXX\_\_PFT.wav.
.3 PROCESS-2\_rec\_\_XXX\_\_PFT.txt.
.3 PROCESS-2\_rec\_\_XXX\_\_CTD.wav.
.3 PROCESS-2\_rec\_\_XXX\_\_CTD.txt.
}

where \texttt{XXX} denotes an anonymised participant identifier.

Each participant contributes six files corresponding to three speech elicitation tasks: SFT, PFT, and CTD.

The entire released dataset (Table \ref{tab:audio_overview}) contains: 
\begin{itemize}
\item 400 participants,
\item 1,200 audio recordings (.wav),
\item 1,200 manual transcript files (.txt),
\item one metadata table (\texttt{meta-info.csv}),
\end{itemize}

\begin{table}[h]
\centering
\caption{Audio characteristics of the PROCESS-2 dataset across elicitation tasks and diagnostic groups. Duration and signal-to-noise ratio (SNR) are reported as mean $\pm$ standard deviation.}
\label{tab:audio_overview}

\setlength{\tabcolsep}{4pt} 

\begin{tabular}{llcccl}
\toprule
\textbf{Task} & \textbf{Diagnosis} & \textbf{N} & \textbf{Duration (s)} & \textbf{SNR (dB)} \\
\midrule

\multirow{3}{*}{SFT} 
 & Dementia & 50  & $59.62 \pm 3.64$ & $-17.01 \pm 6.06$ \\
 & MCI      & 150 & $59.47 \pm 3.95$ & $-17.51 \pm 5.01$ \\
 & HC       & 200 & $59.87 \pm 4.83$ & $-18.05 \pm 5.25$ \\

\midrule

\multirow{3}{*}{PFT} 
 & Dementia & 50  & $60.01 \pm 3.76$ & $-17.01 \pm 6.06$ \\
 & MCI      & 150 & $59.68 \pm 3.66$ & $-17.51 \pm 5.01$ \\
 & HC       & 200 & $60.10 \pm 2.74$ & $-18.05 \pm 5.25$ \\

\midrule

\multirow{3}{*}{CTD} 
 & Dementia & 50  & $61.66 \pm 32.15$ & $-17.01 \pm 6.06$ \\
 & MCI      & 150 & $70.44 \pm 39.45$ & $-17.51 \pm 5.01$ \\
 & HC       & 200 & $74.97 \pm 37.88$ & $-18.05 \pm 5.25$ \\

\midrule
\textbf{Total} & & \textbf{1200} & \textbf{$62.87 \pm 22.86$} & \textbf{$-17.72 \pm 5.28$} \\

\bottomrule
\end{tabular}
\vspace{-5pt}
\end{table}

\begin{table*}[h]
\centering
\caption{Demographic statistics of the PROCESS-2 dataset. Age and MMSE values are reported as mean $\pm$ standard deviation. Gender is reported as male/female counts with percentages. Participant counts ($N$) for the training and test sets show the diagnostic group proportion. MMSE counts ($n$) denote the number of participants with available cognitive scores with the respective percentage in brackets.}
\label{tab:dataset_demographics}

\begin{tabular}{lllcccccccc}
\toprule
\multirow{2.5}{*}{Split} & \multirow{2.5}{*}{Group} & \multirow{2.5}{*}{Diagnosis} & \multirow{2.5}{*}{$N$} & \multirow{2.5}{*}{Split(\%)} & \multirow{2.5}{*}{Age (years)} & \multicolumn{2}{c}{Gender} & \multicolumn{2}{c}{MMSE} \\
\cmidrule(lr){7-8} \cmidrule(lr){9-10}
 &  &  &  &  &  & M/F & Ratio (\%) & Count ($n$) & Mean $\pm$ SD \\
\midrule

\multirow{3}{*}{Train}
 & \multirow{2}{*}{Case} & Dementia & 40 & 80\%  & $75.00 \pm 8.15$ & 28/12 & (70/30\%) & 33 (82.50\%) & $24.36 \pm 4.20$ \\
 &  & MCI      & 120 & 80\% & $71.30 \pm 8.82$ & 59/61 & (49.2/50.8\%) & 84 (70\%) & $26.74 \pm 2.13$ \\
 & Control & HC        & 160 & 80\% & $72.21 \pm 6.92$ & 74/86 & (46.2/53.8\%) & 20 (12.5\%) & $27.57 \pm 6.38$ \\

\midrule

\multirow{3}{*}{Test}
 & \multirow{2}{*}{Case} & Dementia & 10 & 20\%  & $70.30 \pm 7.65$ & 7/3 & (70/30\%) & 9 (90\%) & $23.44 \pm 6.67$ \\
 &  & MCI      & 30 & 20\%  & $70.47 \pm 8.91$ & 15/15 & (50/50\%) & 24 (80\%) & $26.42 \pm 2.70$ \\
 & Control & HC        & 40 & 20\%  & $73.97 \pm 6.94$ & 19/21 & (47.5/52.5\%) & 4 (10\%) & $29.00 \pm 0.82$ \\

\midrule

\multirow{3}{*}{Total}
 & \multirow{2}{*}{Case} & Dementia & 50 & 100\%  & $74.06 \pm 8.20$ & 35/15 & (70/30\%) & 42 (84\%) & $24.17 \pm 4.75$ \\
 &  & MCI      & 150 & 100\% & $71.13 \pm 8.82$ & 74/76 & (49.3/50.7\%) & 108 (72\%) & $26.67 \pm 2.26$ \\
 & Control & HC        & 200 & 100\% & $72.56 \pm 6.94$ & 93/107 & (46.5/53.5\%) & 24 (12\%) & $27.80 \pm 5.85$ \\

\midrule

\multicolumn{3}{l}{\textbf{Grand Total (All)}} & \textbf{400} & --- & \textbf{72.21 $\pm$ 7.89} & \textbf{202/198} & \textbf{(50.5/49.5\%)} & \textbf{174 (43.5\%)} & \textbf{26.22 $\pm$ 3.81} \\

\bottomrule
\end{tabular}
\end{table*}

\subsection{Audio Formats}




The PROCESS-2 dataset recordings are distributed in waveform audio format (.wav), with the corresponding transcriptions provided as UTF-8 encoded plain text files that preserve the original conversational structure, 2.45 GB in total. Across the 1,200 total recordings, file sizes remain relatively compact. The SFT and PFT, each consisting of 400 samples, exhibit very similar distributions, with average file sizes of 1.82 MB and 1.83 MB, respectively. In contrast, the CTD shows significantly higher variability; while it averages 2.19 MB, individual file sizes range from a minimum of 0.26 MB to a maximum of 7.95 MB, likely reflecting the diverse length and detail of participant responses during the picture description task.

\subsection{Metadata Table}

Participant-level metadata are provided in \texttt{meta-info.csv}.  
The table contains the following variables:

\begin{itemize}
\item \textbf{IDs}: anonymised participant directory identifier,
\item \textbf{diagnosis}: clinical diagnostic category (Dementia, MCI, HC),
\item \textbf{age}: participant age at assessment (years),
\item \textbf{gender}: self-reported gender,
\item \textbf{MMSE}: Mini-Mental State Examination score where available,
\item \textbf{Split}: predefined experimental partition (Train/Test).
\end{itemize}

The metadata file enables direct linkage between speech recordings and demographic or clinical characteristics, facilitating reproducible downstream analyses.
All files follow consistent naming conventions to support automated dataset parsing, machine learning benchmarking, and large-scale computational experimentation.

The released dataset represents PROCESS-2 version 1.0. Future updates, corrections, or extensions will be documented through versioned releases to ensure reproducibility of published experiments.

\section{Data Overview} 

Table~\ref{tab:dataset_demographics} summarises the demographic characteristics of the PROCESS-2 dataset across the training and test partitions. 
The dataset contains a total of 400 participants, including 50 participants diagnosed with dementia, 150 individuals with MCI, and 200 HCs. 
Participant age distributions were broadly comparable across dataset splits, with mean ages ranging between 70 and 75 years across diagnostic groups. 
As expected, cognitive scores measured using the MMSE show a decreasing trend from HC to dementia participants. 
Gender distributions were relatively balanced for the HC and MCI groups, while the dementia group contained a higher proportion of male participants. 


Our dataset is imbalanced toward cognitively healthy adults due to open volunteer recruitment; however, this distribution reflects the true prevalence of cognitive impairment in the UK \cite{RCPsych2022}.

Acoustic and linguistic embedding representations were derived using pretrained self-supervised speech models and sentence-level language models to facilitate dataset validation and exploratory analyses. These derived representations are provided as supplementary resources and are primarily used for technical validation rather than constituting core dataset contents.

\section{Technical Validation}

The technical validation of PROCESS-2 aims to verify data reliability, clinical consistency, recording quality, and suitability for computational modelling. 
The PROCESS-2 dataset underwent validation across five key dimensions: demographic integrity, clinical validity, recording and acquisition stability, computational representation analysis, and baseline benchmarking performance.



For continuous variables such as age and MMSE, the normality of the age distributions within each group was first assessed using the Shapiro--Wilk test \cite{royston1992approximating, razali2011power}.
If the normality assumption is violated in at least one diagnostic group, non-parametric statistical tests, such as both a one-way analysis of variance (ANOVA) \cite{fisher1934statistical}, and the Kruskal-Wallis test \cite{kruskal1952use, hecke2012power}, were considered in addition to parametric methods when evaluating group differences. 
To further investigate pairwise differences where the Kruskal-Wallis test produces a significant $p$, Dunn’s post-hoc test with Bonferroni correction \cite{dunn1964multiple, holm1979simple} was applied. 
Pearson correlation coefficients were computed to discover relationships between variables 
\cite{pearson1895vii}. 
All statistical results are summarised in Table~\ref{tab:combined_process2_statistics} and Table \ref{tab:gender_statistics}.

Furthermore, group comparability between training and test subsets was analysed
across diagnostic groups (Dementia, MCI, and HC) for the training and test subsets using raincloud plots. This visualisation combines kernel density estimates, boxplots, and individual observations to provide a detailed view of distributional characteristics and sample variability. To evaluate potential differences between the training and test subsets within each diagnostic group, two-sided Mann-Whitney $U$ tests \cite{mann1947test} were conducted.

\begin{table*}[!ht]
\centering
\caption{Comprehensive statistical evaluation of participant demographics, MMSE, and task-specific metrics. Normality was assessed via Shapiro--Wilk. Group differences were evaluated using one-way ANOVA or Kruskal-Wallis tests. Pairwise comparisons were performed using Dunn’s post-hoc test with Bonferroni correction. Significance levels: *** $p < 0.001$, ** $p < 0.01$, * $p < 0.05$, \textit{ns} $p \geq 0.05$.}
\label{tab:combined_process2_statistics}

\begin{tabular}{lllllcc}
\toprule
\textbf{Task} & \textbf{Variable} & \textbf{Transcription} & \textbf{Analysis} & \textbf{Comparison / Group} & \textbf{Statistic} & \textbf{$p$-value} \\
\midrule

\multicolumn{7}{l}{\textit{Normality Test}} \\
All & Age & -- & Shapiro--Wilk & Dementia & $W=0.973$ & 0.308 (\textit{ns}) \\
All & Age & -- & Shapiro--Wilk & MCI & $W=0.987$ & 0.178 (\textit{ns}) \\
All & Age & -- & Shapiro--Wilk & HC & $W=0.949$ & $1.49\times10^{-6}$ (***) \\
All & MMSE & -- & Shapiro--Wilk & Dementia & $W=0.864$ & $1.45\times10^{-4}$ (***) \\
All & MMSE & -- & Shapiro--Wilk & MCI & $W=0.879$ & $6.47\times10^{-8}$ (***) \\
All & MMSE & -- & Shapiro--Wilk & HC & $W=0.860$ & $3.42\times10^{-3}$ (**) \\

\midrule

\multicolumn{7}{l}{\textit{Global Group Comparison Tests}} \\
All & Age & -- & One-way ANOVA & All groups & $F=3.00$ & 0.05 (*) \\
All & Age & -- & Kruskal--Wallis & All groups & $H=5.17$ & 0.08 (\textit{ns}) \\
All & MMSE & -- & One-way ANOVA & All groups & $F=21.46$ & $4.84\times10^{-9}$ (***) \\
All & MMSE & -- & Kruskal--Wallis & All groups & $H=40.15$ & $1.91\times10^{-9}$ (***) \\

\midrule

\multicolumn{7}{l}{\textit{Pearson Correlations}} \\
All & Age vs MMSE & -- & Pearson $r$ & -- & $r=-0.29$ & -- \\
All & Age vs Diagnosis & -- & Pearson $r$ & -- & $r=0.01$ & -- \\
All & MMSE vs Diagnosis & -- & Pearson $r$ & -- & $r=-0.45$ & -- \\

\midrule

\multicolumn{7}{l}{\textit{Task-Specific Metrics (Duration \& SNR)}} \\
PFT & Duration & -- & Kruskal--Wallis & All groups & $H=1.11$ & 0.574 (\textit{ns}) \\
PFT & SNR & -- & Kruskal--Wallis & All groups & $H=2.20$ & 0.333 (\textit{ns}) \\
CTD & Duration & -- & Kruskal--Wallis & All groups & $H=5.78$ & 0.056 (\textit{ns}) \\
CTD & SNR & -- & Kruskal--Wallis & All groups & $H=2.20$ & 0.333 (\textit{ns}) \\
SFT & Duration & -- & Kruskal--Wallis & All groups & $H=3.20$ & 0.202 (\textit{ns}) \\
SFT & SNR & -- & Kruskal--Wallis & All groups & $H=2.20$ & 0.333 (\textit{ns}) \\

\midrule

\multicolumn{7}{l}{\textit{Embedding-Space Geometry (Distance to HC Centroid)}} \\
SFT & Dist. to HC & Original audio & Kruskal--Wallis & All groups & $H=4.37$ & 0.113 (\textit{ns}) \\
SFT & Dist. to HC & Manual & Kruskal--Wallis & All groups & $H=16.36$ & $2.80\times10^{-4}$ (***) \\
PFT & Dist. to HC & Original audio & Kruskal--Wallis & All groups & $H=3.55$ & 0.169 (\textit{ns}) \\
PFT & Dist. to HC & Manual & Kruskal--Wallis & All groups & $H=3.63$ & 0.163 (\textit{ns}) \\
CTD & Dist. to HC & Original audio & Kruskal--Wallis & All groups & $H=1.52$ & 0.467 (\textit{ns}) \\
CTD & Dist. to HC & Manual & Kruskal--Wallis & All groups & $H=12.05$ & $2.42\times10^{-3}$ (**) \\

\midrule

\multicolumn{7}{l}{\textit{Pairwise Comparisons (Bonferroni Corrected)}} \\
All & MMSE & -- & Dunn Post-hoc & Dementia vs MCI & -- & 0.022 (*) \\
All & MMSE & -- & Dunn Post-hoc & Dementia vs HC & -- & $8.72\times10^{-10}$ (***) \\
All & MMSE & -- & Dunn Post-hoc & MCI vs HC & -- & $1.79\times10^{-6}$ (***) \\
SFT & Dist. to HC & Manual & Dunn Post-hoc & Dementia vs HC & -- & 0.0069 (**) \\
SFT & Dist. to HC & Manual & Dunn Post-hoc & MCI vs HC & -- & 0.0018 (**) \\
CTD & Dist. to HC & Manual & Dunn Post-hoc & Dementia vs HC & -- & 0.0041 (**) \\
CTD & Dist. to HC & Manual & Dunn Post-hoc & MCI vs HC & -- & 0.078 (\textit{ns}) \\

\bottomrule
\end{tabular}
\vspace{-0pt}
\end{table*}

\begin{table*}[!ht]
\centering
\caption{Statistical evaluation of demographic variables in the PROCESS-2 dataset. Gender distributions were compared using chi-square tests of independence. Effect size is reported using Cramér’s $V$.}
\label{tab:gender_statistics}

\begin{tabular}{llcccc}
\toprule
Comparison & Test & Statistic & $p$-value & Effect size & Interpretation \\
\midrule
Diagnosis vs Gender & $\chi^2(2)$ & 8.97 & 0.011 & $V = 0.15$ & Small effect \\
Train vs Test & $\chi^2(1)$ & 0.00 & 0.98 & -- & No difference \\
\bottomrule
\end{tabular}

\end{table*}

\subsection{Demographic and Clinical Cohort Validation}

Demographic distributions were analysed to determine whether diagnostic categories differed systematically in age, gender composition, or cognitive severity (MMSE).

\subsubsection{Age}

Table~\ref{tab:combined_process2_statistics} indicated that the HC group significantly deviated from a normal distribution ($W=0.949$, $p<0.001$), whereas the MCI ($W=0.987$, $p=0.178$) and dementia ($W=0.973$, $p=0.308$) groups did not exhibit significant deviations from normality. 
Because the normality assumption was violated in at least one diagnostic group, non-parametric statistical tests were considered in addition to parametric methods when evaluating group differences.

To assess whether mean age differed across diagnostic categories, both a one-way analysis of variance (ANOVA) and the non-parametric Kruskal-Wallis test were conducted. 
ANOVA yielded $F=3.00$ ($p=0.05$), while the Kruskal--Wallis test showed no significant difference ($H=5.17$, $p=0.08$). The convergence of these tests indicates that diagnostic groups are age-comparable, reducing the likelihood of age acting as a confounding factor at the conventional significance level ($\alpha=0.05$) in downstream analyses.

To further investigate potential pairwise differences between diagnostic groups, Dunn’s post-hoc test with Bonferroni correction was applied. The adjusted $p$-values for all pairwise comparisons were greater than $0.05$, including dementia vs. MCI ($p=0.101$), dementia vs. HC ($p=0.771$), and MCI vs. HC ($p=0.361$). These findings confirm that no statistically significant age differences exist between the diagnostic groups.

Overall, the statistical analysis carried out in Table \ref{tab:combined_process2_statistics} and Figure \ref{fig:age_dist} shows that the age distributions are broadly comparable across diagnostic categories in the PROCESS-2 dataset, reducing the likelihood that age acts as a confounding factor in downstream analyses of speech-based cognitive decline detection.
The training and test partitions also exhibit broadly comparable age characteristics across diagnostic groups.

\subsubsection{Gender}

Table~\ref{tab:gender_statistics} summarises the statistical evaluation of gender distributions in the PROCESS-2 dataset. 
The analysis revealed a statistically significant association between gender and diagnosis ($\chi^2(2)=8.97$, $p=0.011$). Inspection of the observed and expected frequencies indicated a higher proportion of male participants in the dementia group and a higher proportion of female participants in the HC group. However, the effect size was small (Cramér’s $V=0.15$), suggesting that the magnitude of this association is weak.

In contrast, no significant difference in gender distribution was observed between the training and test subsets ($\chi^2 \approx 0.00$, $p=0.98$), indicating that the dataset partitions are well balanced with respect to gender (Table \ref{tab:combined_process2_statistics} and Figure \ref{fig:gender_dist}).

\subsubsection{MMSE}

Table \ref{tab:combined_process2_statistics} indicated that all groups, as expected, significantly deviated from normality, including dementia ($W=0.864$, $p<0.001$), MCI ($W=0.879$, $p<0.001$), and HC ($W=0.860$, $p<0.01$). 
Although a one-way ANOVA also indicated a significant effect of diagnosis on MMSE scores ($p<0.001$), the violation of normality assumptions and the bounded nature of MMSE scores make parametric results less reliable. Therefore, the non-parametric Kruskal-Wallis test is considered more appropriate and is used for interpretation, and it indicates that cognitive performance differs across diagnostic categories.

To further investigate pairwise differences, Dunn’s post-hoc test with Bonferroni correction \cite{dunn1964multiple, holm1979simple} was applied. The results showed significant differences between all diagnostic groups, including dementia vs. MCI ($p=0.022$), dementia vs. HC ($p<0.001$), and MCI vs. HC ($p<0.001$). The biggest differences were observed between dementia and HC, and between MCI and HC, while the difference between dementia and MCI, although statistically significant, was comparatively smaller.

\begin{figure*}[!ht]
    \centering
    
    \begin{subfigure}{\textwidth}
        \includegraphics[width=\textwidth]{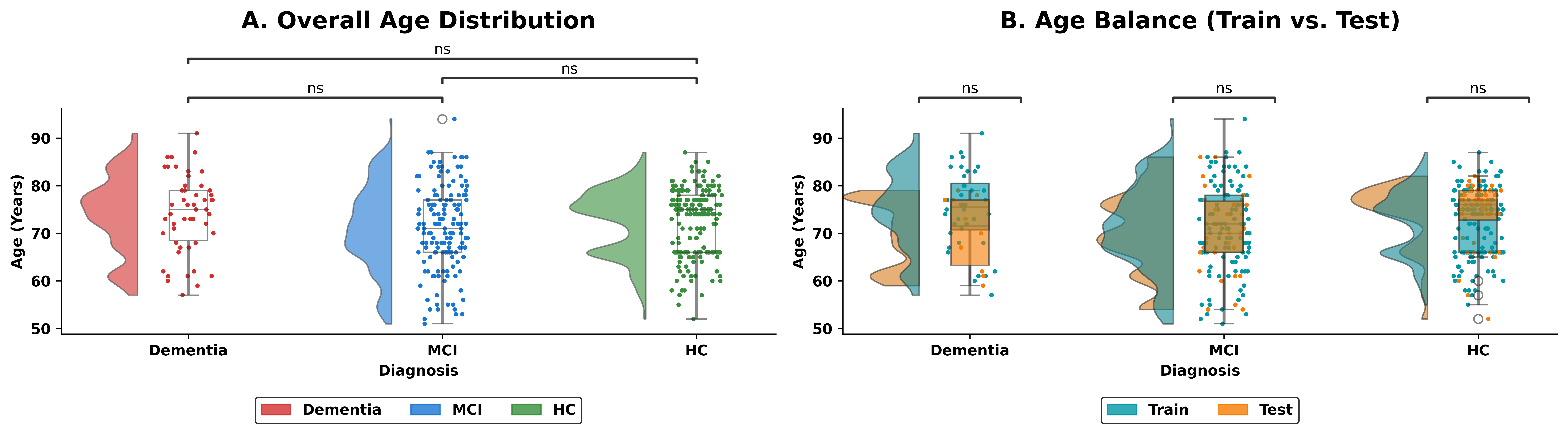}
        \caption{Age distribution}
        \label{fig:age_dist}
    \end{subfigure}
    
    \vspace{0.5em}
    
    \begin{subfigure}{\textwidth}
        \includegraphics[width=\textwidth]{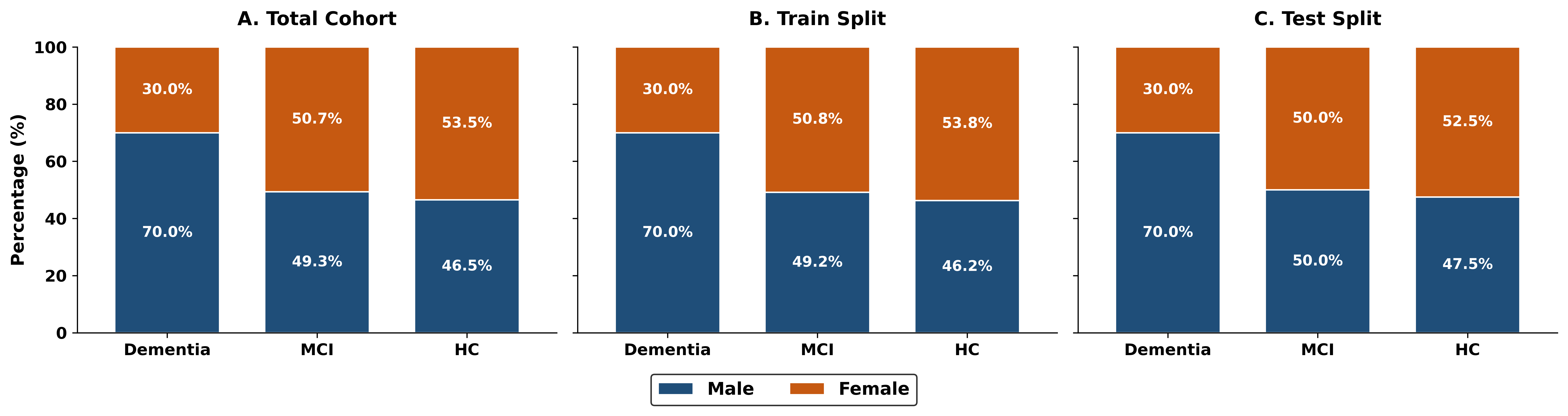}
        \caption{Gender distribution}
        \label{fig:gender_dist}
    \end{subfigure}
    
    \vspace{0.5em}
    
    \begin{subfigure}{\textwidth}
        \includegraphics[width=\textwidth]{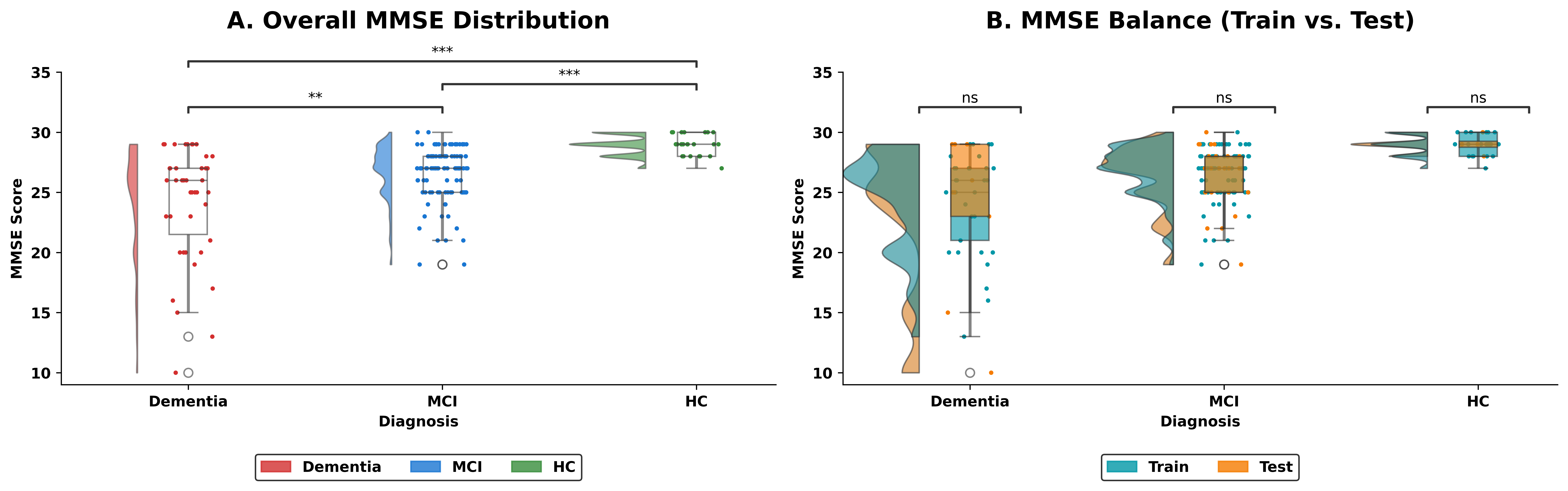}
        \caption{MMSE distribution}
        \label{fig:mmse_dist}
    \end{subfigure}

    \caption{\textbf{Demographic and clinical characterisation of the PROCESS-2 dataset.} 
    \textbf{(a) Top (Age):} Raincloud plots showing age distribution across diagnostic groups (Dementia, MCI, HC) and data splits (Train/Test). 
    \textbf{(b) Middle (Gender):} Stacked bar plots representing male (navy) and female (terracotta) distribution for the total cohort and splits. 
    \textbf{(c) Bottom (MMSE):} Distribution of Mini-Mental State Examination scores illustrating clinical progression and parity between Train (cyan) and Test (orange) subsets. 
    Statistical significance for age and MMSE was determined using the Kruskal-Wallis test with Bonferroni correction ($^{***}p < 0.001$, $^{**}p < 0.01$, \textit{ns}: non-significant). Individual data points, boxplots, and probability density distributions provide a comprehensive view of data variance and gender ratios.}
    
    \label{fig:combined_demographics}
    \vspace{-15pt}
\end{figure*}

Overall, these findings confirm that MMSE scores provide clear separation between diagnostic groups, supporting the clinical validity of the dataset and indicating that cognitive status is strongly reflected in the recorded measures.
However, the distributions of MMSE do not differ among train and test splits for every diagnosis, as demonstrated in Figure \ref{fig:mmse_dist}.

\subsubsection{Correlation structure.}

Table~\ref{tab:combined_process2_statistics} also illustrates correlations between selected metadata variables in the PROCESS-2 dataset. Age exhibited a weak negative correlation with MMSE scores ($r=-0.29$), suggesting slightly lower cognitive scores in older participants. A moderate negative correlation was observed between MMSE scores and diagnostic labels ($r=-0.45$), reflecting the expected clinical relationship whereby participants with greater cognitive impairment tend to exhibit lower MMSE scores. In contrast, age showed negligible correlation with diagnostic category ($r=0.01$), indicating that diagnostic groups are not strongly confounded by age within the dataset.

The weak correlation between age and diagnosis confirms that diagnostic labels primarily reflect cognitive status rather than demographic bias.

    


\subsection{Audio Acquisition Consistency}


Recording stability was assessed using task duration and signal-to-noise ratio (SNR), summarised in Table~\ref{tab:audio_overview}. 
Duration statistics reveal greater variability in the CTD task, reflected by higher standard deviations compared to the fluency tasks. This is expected given its open-ended, spontaneous nature, whereas PFT and SFT exhibit tightly controlled durations due to fixed time constraints.
Statistical analysis (Table~\ref{tab:combined_process2_statistics}) showed no significant differences in duration across diagnostic groups for any task (PFT: $p=0.574$, SFT: $p=0.202$, CTD: $p=0.056$), although CTD approached significance, with a marginal trend between dementia and HC.

\begin{figure*}[!ht]
    \centering

    \includegraphics[width=\textwidth]{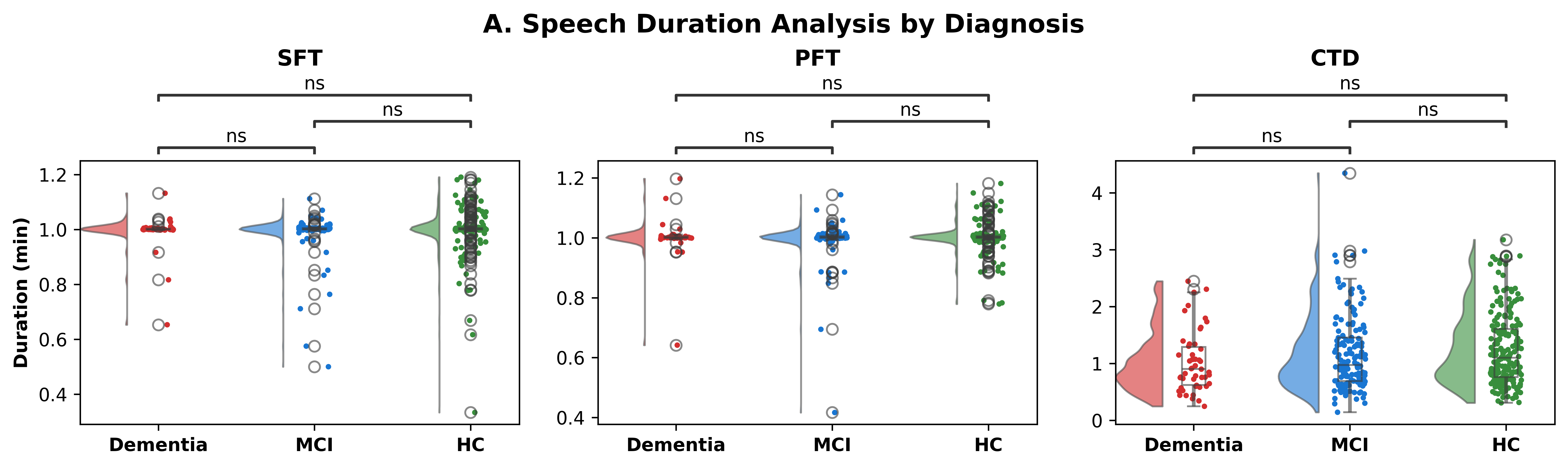}

    \includegraphics[width=\textwidth]{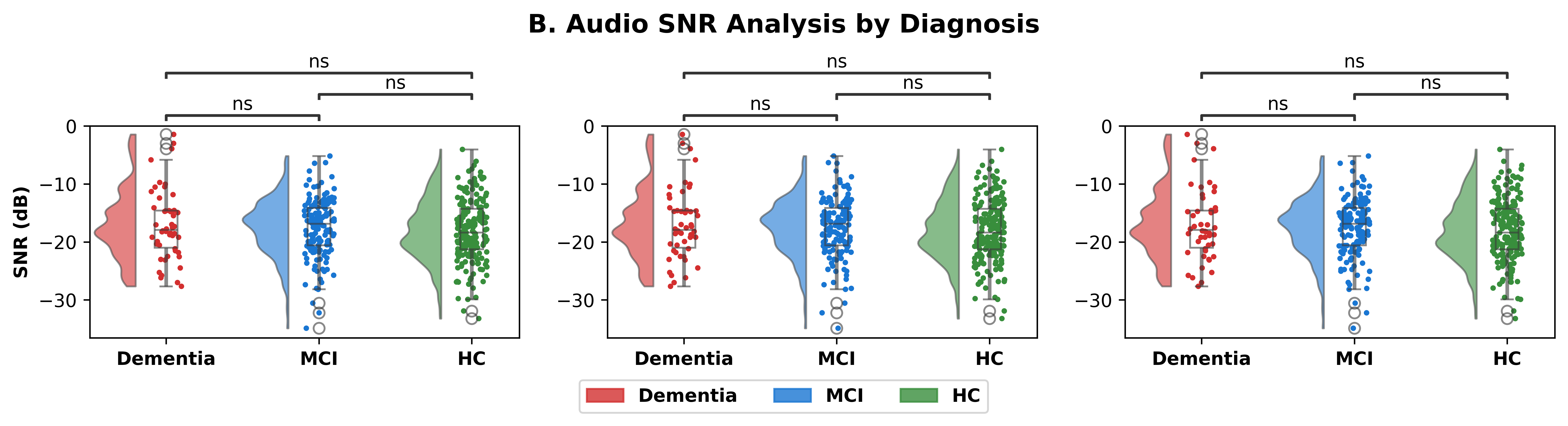}
    \vspace{0.2em}

    \includegraphics[width=\textwidth]{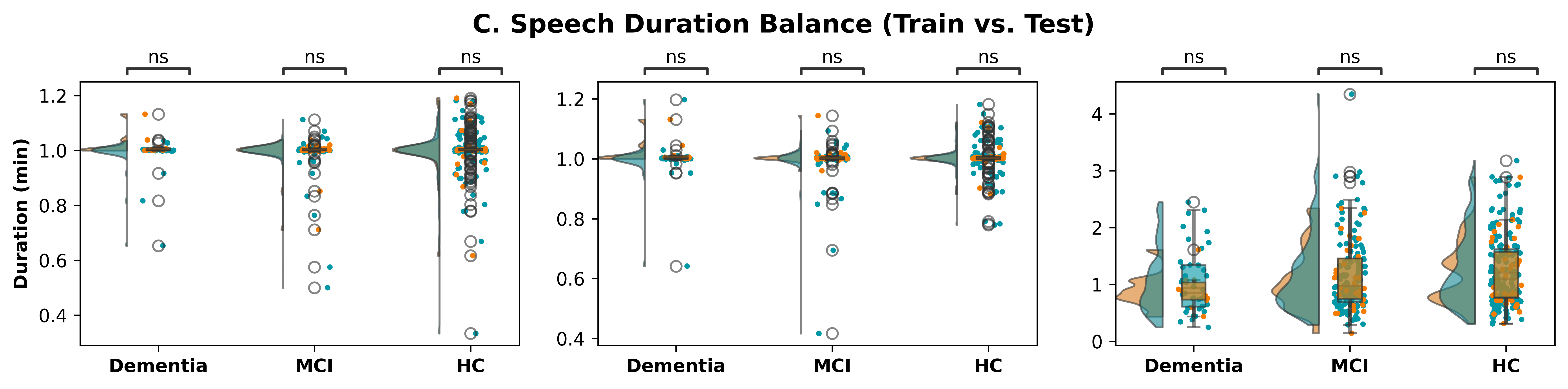}

    \includegraphics[width=\textwidth]{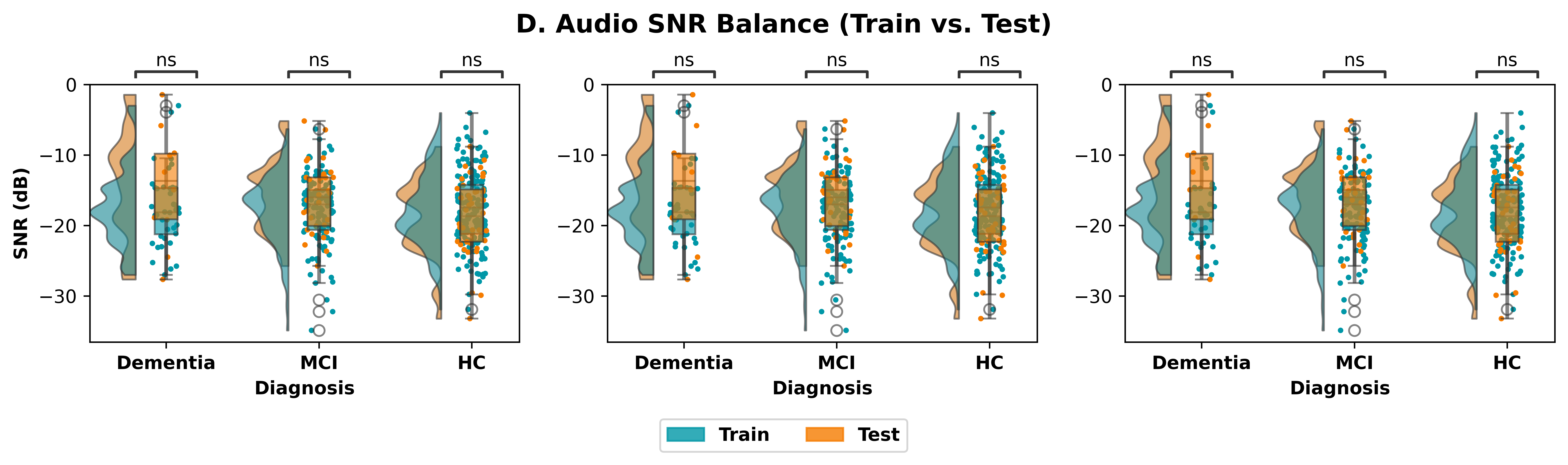}

    \caption{\textbf{Audio data characterisation across diagnostic groups and dataset splits.}
    \textbf{(A) Speech Duration by Diagnosis:} Raincloud plots showing the distribution of speech duration (minutes) across diagnostic groups (Dementia, MCI, HC) for each speech task.
    \textbf{(B) Audio Quality (SNR) by Diagnosis:} Signal-to-noise ratio (dB) distributions across diagnostic groups, highlighting recording quality consistency.
    \textbf{(C) Speech Duration Balance (Train vs. Test):} Comparison of duration distributions across diagnostic groups stratified by data split, demonstrating balance between training and test sets.
    \textbf{(D) Audio Quality Balance (Train vs. Test):} SNR distributions across splits, confirming comparable recording conditions between training and evaluation subsets.
    Statistical significance was assessed using the Kruskal–Wallis test with Bonferroni correction ($^{***}p < 0.001$, $^{**}p < 0.01$, \textit{ns}: non-significant). Raincloud plots combine raw data points, boxplots, and kernel density estimates to provide a comprehensive view of distributional characteristics.}
    
    \label{fig:audio_characterization}
\end{figure*}

To evaluate audio quality, we calculated the SNR between speech and pauses by identifying speech segments using Silero VAD, a highly accurate, lightweight and efficient voice activity detector \cite{Silero_VAD}. 
SNR values were consistent across tasks and diagnostic groups (approximately $-17$ to $-18$ dB), with no significant group differences ($p=0.333$), indicating stable recording conditions and preprocessing.
As illustrated in Figure~\ref{fig:audio_characterization}, both duration and SNR distributions remain comparable across diagnostic groups and between training and test splits, confirming the absence of systematic biases.
Overall, these results demonstrate that while task design influences duration variability, the dataset remains well-balanced in terms of recording structure and audio quality, reducing the likelihood of confounding effects in downstream analyses.

\subsection{Embedding-space geometric analysis.}

\begin{figure*}[!ht]
    \centering
    \begin{subfigure}{\textwidth}
        \centering
        \includegraphics[width=\linewidth]{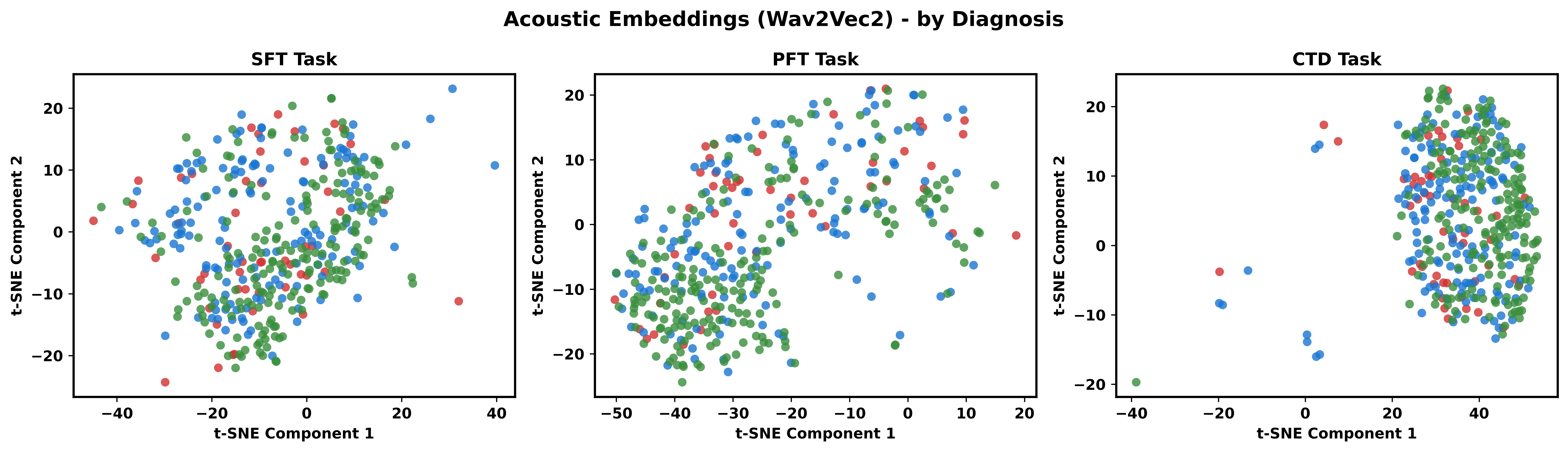}
        \label{fig:acoustic_diag}
    \end{subfigure}
    \begin{subfigure}{\textwidth}
        \centering
        \includegraphics[width=\linewidth]{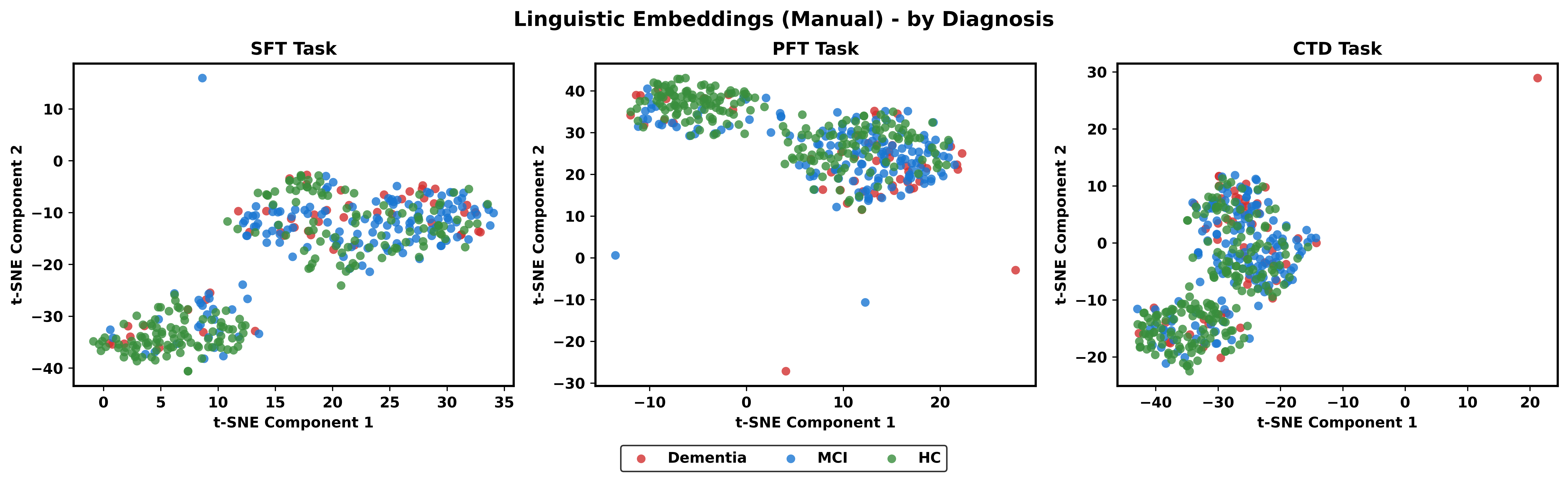}
        \label{fig:ling_diag}
    \end{subfigure}
    \begin{subfigure}{\textwidth}
        \centering
        \includegraphics[width=\linewidth]{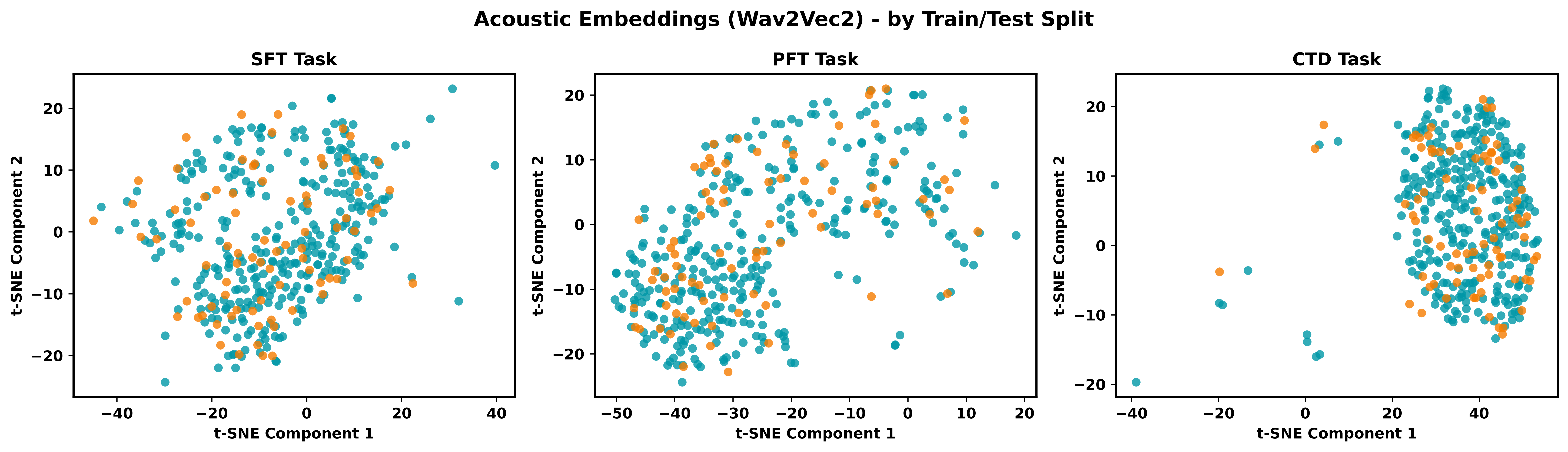}
        \label{fig:acoustic_split}
    \end{subfigure}
    \begin{subfigure}{\textwidth}
        \centering
        \includegraphics[width=\linewidth]{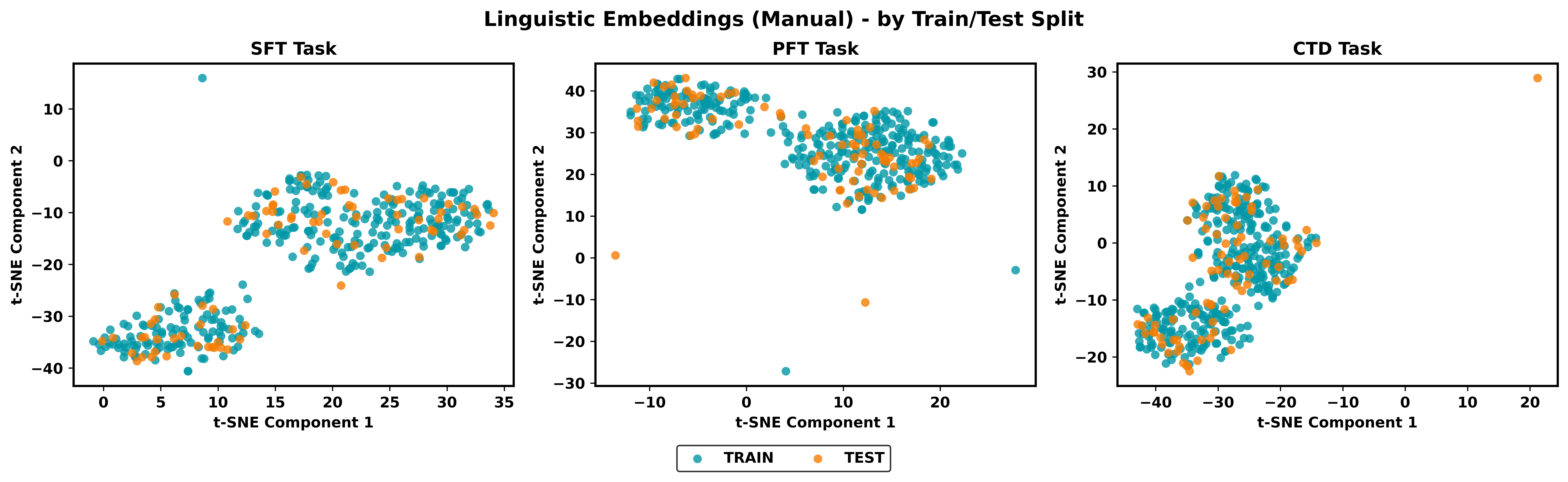}
        \label{fig:ling_split}
    \end{subfigure}
    \caption{\textbf{t-SNE visualisation of acoustic and linguistic embeddings across tasks, diagnostic groups, and dataset splits.}
    (Top) Acoustic embeddings (\WV{}) coloured by diagnosis.
    (Second) Linguistic embeddings (manual transcripts) coloured by diagnosis.
    (Third) Acoustic embeddings coloured by Train/Test split.
    (Bottom) Linguistic embeddings coloured by train/test split.
    Each point represents a recording projected into two dimensions, illustrating clustering patterns and consistency across tasks and splits.}
    \label{fig:overall_embeddings}
    \vspace{-15 pt}
\end{figure*}

\begin{table*}[!ht]
\centering
\small
\setlength{\tabcolsep}{4pt}
\caption{Comparison of classical models (LR and MLP) and LLMs (\distilbert{} and \roberta{}) for 2-way (2w) and 3-way (3w) classification (Macro $F_1$) and regression (RMSE) using both acoustic (Acous.) and linguistic (Ling.) features. Best results per category and task are in bold.}
\label{tab:merged_results}
\begin{tabular}{lllcccccccccccc}
\toprule
 &  &  & \multicolumn{6}{c}{Classical} & \multicolumn{6}{c}{LLMs} \\
\cmidrule(lr){4-9} \cmidrule(lr){10-15}

 &  &  & \multicolumn{4}{c}{Macro $F_1$} & \multicolumn{2}{c}{RMSE} 
   & \multicolumn{4}{c}{Macro $F_1$} & \multicolumn{2}{c}{RMSE} \\

\cmidrule(lr){4-7} \cmidrule(lr){8-9}
\cmidrule(lr){10-13} \cmidrule(lr){14-15}

 &  &  & \multicolumn{2}{c}{LR} & \multicolumn{2}{c}{MLP} 
   & DT & RF
   & \multicolumn{2}{c}{\distilbert{}} & \multicolumn{2}{c}{\roberta{}}
   & \distilbert{} & \roberta{} \\

\cmidrule(lr){4-5} \cmidrule(lr){6-7}
\cmidrule(lr){10-11} \cmidrule(lr){12-13}

Task & Type & Representation 
& 2w & 3w & 2w & 3w & Reg & Reg 
& 2w & 3w & 2w & 3w & Reg & Reg \\

\midrule
\multirow{4}{*}{SFT}
& Acous. & \compare{} & 0.62 & 0.39 & 0.63 & 0.42 & 4.18 & 4.12 & - & - & - & - & - & - \\
\cmidrule{2-15}
& \multirow{3}{*}{Ling.}
& Manual & 0.74 & 0.43 & 0.71 & 0.47 & 5.38 & 4.32 & \textbf{0.82} & \textbf{0.59} & 0.79 & 0.56 & 4.11 & \textbf{4.10} \\
& & ASR (\WV{}) & 0.59 & 0.38 & 0.55 & 0.31 & 5.29 & 4.39 & 0.71 & 0.43 & 0.68 & 0.37 & 4.11 & 4.11 \\
& & ASR (\Whisper{}) & 0.60 & 0.35 & 0.58 & 0.41 & 4.82 & 4.15 & 0.71 & 0.46 & 0.71 & 0.47 & 4.11 & 4.11 \\

\midrule
\multirow{4}{*}{PFT}
& Acous. & ComParE & 0.61 & 0.47 & 0.60 & 0.44 & 4.86 & 4.19 & - & - & - & - & - & - \\
\cmidrule{2-15}
& \multirow{3}{*}{Ling.}
& Manual & 0.72 & 0.51 & 0.73 & 0.55 & 4.13 & 4.18 & 0.80 & 0.55 & \textbf{0.81} & \textbf{0.57} & 4.10 & 4.13 \\
& & ASR (\WV{}) & 0.58 & 0.42 & 0.62 & 0.38 & 4.77 & 4.14 & 0.71 & 0.53 & 0.75 & 0.51 & 4.11 & 4.11 \\
& & ASR (\Whisper{}) & 0.52 & 0.29 & 0.57 & 0.35 & 4.14 & 4.11 & 0.75 & 0.52 & 0.72 & 0.57 & \textbf{4.09} & 4.11 \\

\midrule
\multirow{4}{*}{CTD}
& Acous. & ComParE & 0.60 & 0.39 & 0.70 & 0.48 & 4.33 & 4.11 & - & - & - & - & - & - \\
\cmidrule{2-15}
& \multirow{3}{*}{Ling.}
& Manual & 0.69 & 0.49 & 0.76 & 0.56 & 4.09 & 4.08 & \textbf{0.85} & \textbf{0.58} & 0.82 & 0.55 & 4.10 & 4.13 \\
& & ASR (\WV{}) & 0.59 & 0.40 & 0.66 & 0.38 & 5.00 & 4.19 & 0.72 & 0.51 & 0.70 & 0.37 & \textbf{3.87} & 4.13 \\
& & ASR (\Whisper{}) & 0.65 & 0.48 & 0.62 & 0.44 & 4.61 & 4.04 & 0.64 & 0.47 & 0.66 & 0.43 & 4.09 & 4.13 \\

\midrule
\multirow{4}{*}{Fluency}
& Acous. & ComParE & 0.58 & 0.29 & 0.56 & 0.46 & 5.56 & 4.10 & - & - & - & - & - & - \\
\cmidrule{2-15}
& \multirow{3}{*}{Ling.}
& Manual & 0.64 & 0.38 & 0.69 & 0.42 & 4.64 & 4.16 & 0.81 & \textbf{0.53} & \textbf{0.85} & 0.52 & 4.12 & 4.11 \\
& & ASR (\WV{}) & 0.52 & 0.36 & 0.61 & 0.40 & 4.60 & 4.19 & 0.74 & 0.39 & 0.72 & 0.48 & 4.12 & 4.11 \\
& & ASR (\Whisper{}) & 0.56 & 0.37 & 0.53 & 0.37 & 4.42 & 4.11 & 0.72 & 0.50 & 0.71 & 0.44 & \textbf{4.08} & 4.11 \\

\midrule
\multirow{4}{*}{ALL}
& Acous. & ComParE & 0.62 & 0.49 & 0.60 & 0.26 & 5.01 & 4.12 & - & - & - & - & - & - \\
\cmidrule{2-15}
& \multirow{3}{*}{Ling.}
& Manual & 0.75 & 0.28 & 0.74 & 0.42 & 4.59 & 4.17 & \textbf{0.85} & \textbf{0.48} & 0.81 & 0.46 & 4.13 & \textbf{4.11} \\
& & ASR (\WV{}) & 0.67 & 0.25 & 0.64 & 0.35 & 4.69 & 4.26 & 0.77 & 0.36 & 0.70 & 0.27 & 4.13 & \textbf{4.11} \\
& & ASR (\Whisper{}) & 0.65 & 0.25 & 0.65 & 0.31 & 4.73 & 4.16 & 0.80 & 0.33 & 0.74 & 0.37 & 4.12 & \textbf{4.11} \\

\bottomrule
\end{tabular}
\vspace{-10pt}
\end{table*}

To quantify disease-related structure in the learned representation space, we measured the geometric distance of each participant embedding from the HC group centroid.
Let $\mathbf{x}_i \in \mathbb{R}^{d}$ denote the embedding vector of participant $i$, where $d$ is the embedding dimensionality. For each task and transcription condition, the HC centroid was computed as:
\begin{equation}
\mathbf{c}_{HC} =
\frac{1}{N_{HC}}
\sum_{i \in HC} \mathbf{x}_i ,
\end{equation}
where $N_{HC}$ is the number of HC participants. Disease-related deviation was then quantified as the Euclidean distance:
\begin{equation}
D_i = \lVert \mathbf{x}_i - \mathbf{c}_{HC} \rVert_2 = \sqrt{\sum_{k=1}^{d}(x_{ik}-c_{HC,k})^2}.
\end{equation}
This metric reflects the extent to which an individual’s representation deviates from the normative healthy embedding space.
Distances to the HC centroid revealed task-dependent differences in embedding-space organisation (Table~\ref{tab:combined_process2_statistics}). 
Significant group effects were observed for the SFT and CTD under manual transcription, whereas no significant separation was found for original audio embeddings or the PFT.
Post-hoc analysis showed that, for SFT manual transcriptions, both MCI and dementia groups exhibited significantly greater distances from the HC centroid, indicating progressive deviation of linguistic representations. Similarly, CTD manual transcriptions showed significant separation primarily driven by increased distances in the dementia group relative to HC.
In contrast, original audio embeddings did not yield significant group differences, suggesting that transcription-derived linguistic features capture disease-related variation more effectively than acoustic representations alone. As illustrated in Figure~\ref{fig:overall_embeddings}, embedding distributions also remain consistent between training and test splits across diagnostic groups, indicating the absence of dataset shift.
Overall, disease progression is reflected as increasing geometric displacement from the HC centroid, demonstrating that PROCESS-2 captures clinically meaningful variation in representation space and provides a robust benchmark for representation learning.

\subsection{Benchmark Modelling Experiments}

Benchmark experiments were conducted to evaluate the suitability of PROCESS-2 for automatic cognitive assessment (Table~\ref{tab:merged_results}). We compared classical machine learning models and transformer-based language models across multiple tasks and feature representations using 2-way (200 case vs 200 control) and 3-way (50 dementia vs 150 MCI vs 200 HC) classification and regression (174 participants with MMSE scores) strategies (Table \ref{tab:dataset_demographics}).
Moreover, we combined all three tasks and represented them as the `ALL' task.
Additionally, automatic speech recognition (ASR) transcripts generated using \Whisper{} medium and \WV{} were evaluated alongside manual transcripts to assess model robustness under realistic transcription conditions. The corresponding word error rates (WER) were approximately 40\% and 60\%, respectively. Elevated WER values primarily reflect the intentional preservation of conversational disfluencies and speaker identifiers (Section \ref{subsec:data-curation}), which better approximate real-world clinical deployment scenarios.

Analysis of the experimental results reveals that LLMs consistently outperform classical models, logistic regression (LR) and multilayer perceptrons (MLP), across both classification complexities, particularly when using manual transcripts. 
In 2-way classification, LLMs (\distilbert{} and \roberta{}) achieve a peak Macro $F_1$ of 0.85, significantly outpacing the 0.76 reached by Classical MLP models. 
This performance gap remains present but narrows in the more challenging 3-way classification task, where LLMs reach a top $F_1$ score of 0.59 (SFT task) compared to the Classical peak of 0.56 (CTD task). 
While both architectures struggle more with the 3-way diagnostic split, the LLMs' ability to leverage deep linguistic context provides a superior edge in distinguishing between HC, MCI, and Dementia. 
This effectiveness extends to regression tasks as well, with \distilbert{} achieving the best overall RMSE of 3.87, demonstrating that modern transformer-based models are currently the most robust choice for automated cognitive screening.

Overall, these results indicate that (i) linguistic representations provide stronger predictive signals than acoustic features, (ii) transformer-based models outperform classical approaches, particularly for classification, and (iii) the dataset supports robust modelling even with automatically generated transcripts.
Performance is comparable to established benchmarks such as \adress{} and ADReSSo, confirming that PROCESS-2 constitutes a realistic and challenging evaluation resource. Comparable results between training and test sets further indicate the absence of dataset leakage and validate the predefined split strategy, with all preprocessing and model development performed exclusively on training data.

\begin{figure}[h]
  \centering 
  \includegraphics[width=\linewidth]{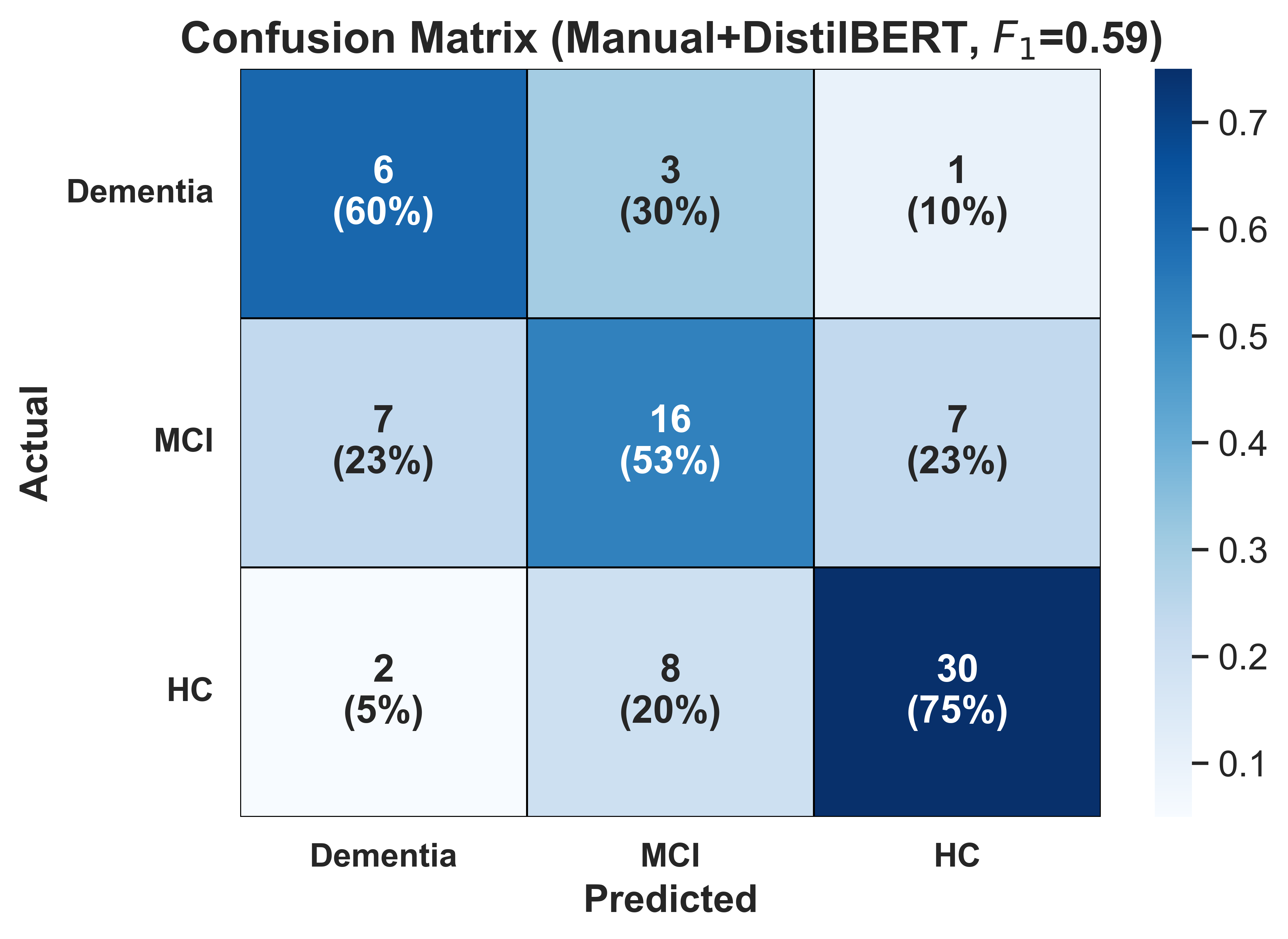} 
  \caption{
    Confusion matrix for three-way diagnostic classification (Dementia, MCI, HC) using manual transcripts and a DistilBERT model ($F_1 = 0.59$ in Table \ref{tab:merged_results}) on SFT. 
    Cells report absolute counts together with row-wise percentages, indicating the proportion of participants within each true diagnostic group assigned to each predicted class. 
    }
  \label{fig:conf-mat}
  \vspace{-10pt}
\end{figure}

Figure \ref{fig:conf-mat} illustrates the confusion matrix for the three-class diagnostic classification task for best-preformed 3-way classifcation, using manual transcripts and a \distilbert{} model using SFT producing Macro $F_1 = 0.59$ in Table \ref{tab:merged_results}. 
Row-wise normalisation highlights classification behaviour within each clinical group. 
Dementia participants were correctly identified in 6 of 10 cases (60\%), while 30\% were predicted as MCI and 10\% as HC. 
MCI cases exhibited greater diagnostic ambiguity, with 16 of 30 participants correctly classified (53\%), and equal proportions misclassified as dementia (23\%) and HC (23\%). 
HC participants were identified with the highest reliability, with 30 of 40 participants correctly classified (75\%), while 20\% were predicted as MCI and only 5\% as dementia.
Importantly, most errors occur between neighbouring diagnostic categories rather than extreme misclassifications (e.g., dementia directly classified as HC), suggesting that the model captures a graded cognitive severity continuum consistent with clinical progression from HC to MCI to dementia.

\subsection{Ecological Validity and Real-World Constraints}

Unlike laboratory corpora, PROCESS-2 intentionally preserves real-world spontaneous conversational characteristics, including pauses, assisting speakers, and environmental variability of speech. Some participants required physical assistance during assessments; these interactions were retained to reflect realistic deployment scenarios for digital cognitive screening tools.

Manual inspection ensured recording completeness. Rare interruptions in time-limited tasks were trimmed to active response segments without altering speech content.

The dataset exhibits high ecological validity, as recordings were collected in participants’ natural environments using heterogeneous consumer devices rather than controlled laboratory settings.
This design enables evaluation of algorithms under deployment conditions that closely resemble real-world digital cognitive assessment settings.

\subsection{Summary of Validation Findings}

Collectively, the validation analyses demonstrate that PROCESS-2 exhibits:

\begin{itemize}
\item demographic comparability across diagnostic groups,
\item clinically meaningful MMSE separation,
\item stable recording quality despite remote acquisition,
\item measurable disease-related structure in representation space,
\item reproducible benchmark modelling performance.
\end{itemize}

These results confirm that PROCESS-2 constitutes a reliable, high-quality dataset suitable for reproducible research in speech-based cognitive assessment.


\section{Data Availability}





The PROCESS-2 dataset contains human speech recordings collected under clinical ethical approval and therefore cannot be released as unrestricted public data. Access is provided through a controlled access framework to ensure responsible reuse and protection of participant privacy.

The dataset is hosted on the Hugging Face data repository:

\begin{center}
\url{https://huggingface.co/datasets/CognoSpeak/PROCESS-2}
\end{center}

Access to the dataset requires submission of a request describing institutional affiliation and intended research use. Applicants must agree to the PROCESS-2 Data Use Agreement prior to access being granted. Approved researchers obtain access through a gated repository mechanism.
The released dataset includes:

\begin{itemize}
\item  speech recordings in waveform audio format (.wav),
\item  manually generated transcripts (.txt),
\item  participant-level metadata tables,
\item  dataset documentation and usage guidelines.
\end{itemize}

All participants provided informed consent permitting controlled research data sharing, and the dataset was anonymised prior to release and contains no direct personal identifiers. 
Redistribution, commercial use, or attempts at participant re-identification are prohibited under the data use agreement.
Due to the inherently identifiable nature of human voice recordings, 
fully open public release of raw audio is not ethically appropriate. 
Although all recordings were anonymised, speech signals themselves 
may retain biometric characteristics. Therefore, access to the dataset 
is provided under controlled conditions through a data access agreement 
to protect participant privacy while enabling legitimate research use.
Researchers using the PROCESS-2 dataset must cite this data descriptor.



\section{Code Availability}

All code required to reproduce statistical analyses, embedding 
generation, and baseline modelling experiments along with associated Anaconda environments are publicly available at:

\begin{center}
\url{https://github.com/CognoSpeak/PROCESS-2}
\end{center}


The codebase is released under the Apache License 2.0, permitting reuse, 
modification, and redistribution subject to the terms of the licence. 
Version control is maintained through GitHub to ensure transparency and 
reproducibility of all results reported in this study.
Code archived at Zenodo DOI:
\begin{center}
\url{https://doi.org/10.5281/zenodo.19900225}
\end{center}

All analyses reported in this study can be reproduced using the publicly available code \cite{pahar_2026_PROCESS_codes} together with approved access to the PROCESS-2 dataset. 

\bibliographystyle{IEEEtran}
\bibliography{mybib}

\section{Author Contributions}
Madhurananda Pahar conceived the study, designed and conducted the experiments, performed data analysis, classification, and authored the published Python code and manuscript. 
Bahman Mirheidari contributed to experimental implementation, provided technical guidance, and assisted with manuscript revision. 
Hend Elghazaly, Fritz Peters, Sophie Young, Labhpreet Kaur, Caitlin Illingworth, and Wing-Zin Leung provided assistance in checking the audio data and preparing the transcripts. 
Caitlin Illingworth and Daniel Blackburn led participant recruitment and data collection across the community centres and assisted in writing the Methods section. 
Heidi Christensen supervised the research, contributed to the study design and interpretation of results, and provided overall project guidance.

All authors contributed to reviewing and editing the manuscript and approved the final version.

\section{Competing Interests}
The authors declare no competing interests.

\section{Acknowledgements} 
We acknowledge the support of NHS clinicians, who recruited the participants, and Therapy Box, with whom we co-developed the data collection front-end app for \cognospeak{}, formerly \cognospeakOLD{} assessments.
We also acknowledge Jon Barker and Robbie Sutherland for assisting in publishing the data through the Hugging Face portal and preparing the data-release terms and conditions. 
The views expressed are those of the authors and not necessarily those of the NHS, the NIHR or the Department of Health and Social Care (DHSC).
For the purpose of open access, the author has applied a Creative Commons Attribution (CC BY) licence to any Author Accepted Manuscript version arising. 

\section{Funding}
This research was partly funded by the NIHR Sheffield Biomedical Research Centre (BRC), and the NIHR202911 award under the NIHR i4i programme.

\end{document}